\documentclass[reprint,
nofootinbib,
 amsmath,amssymb,
 aps,
 amsfonts,
]{revtex4-2}

\usepackage{graphicx}
\usepackage{dcolumn}
\usepackage{bm}
\usepackage{braket}
\usepackage{euscript}
\usepackage{xcolor}
\usepackage{nicefrac}
\usepackage{comment}
\usepackage{siunitx}
\usepackage[caption=false]{subfig}

\begin{document}


\title{Trap induced broadening in a potential hydrogen lattice clock}

\author{J. P. Scott}
\email{joseph.p.scott@durham.ac.uk}
\affiliation{%
 Department of Physics, Durham University, Durham, United Kingdom\\
}%

\author{R. M. Potvliege}
\affiliation{%
 Department of Physics, Durham University, Durham, United Kingdom\\
}%

\author{D. Carty}
\affiliation{%
 Department of Physics, Durham University, Durham, United Kingdom\\
}%

\author{M. P. A. Jones}
\affiliation{%
 Department of Physics, Durham University, Durham, United Kingdom\\
}%

\date{\today}

\begin{abstract}
We consider the potential use of optical traps for precision measurements in atomic hydrogen (H). 
Using an implicit summation method, we calculate the atomic polarisability, the rates of elastic/inelastic scattering and the ionisation rate in the wavelength range 395 to 1000\,nm. 
We extend previous work to predict three new magic wavelengths for the 1S--2S transition. 
At the magic wavelengths, the 1S--2S transition is unavoidably and significantly broadened due to trap-induced ionisation associated with the high intensity required to trap the 1S state.
However, we also find that this effect is partially mitigated by the low mass of H, which increases the trap frequency, enabling Lamb-Dicke confinement in shallow lattices. We find that a H optical lattice clock, free from the motional systematics which dominate in beam experiments, could operate with an intrinsic linewidth of O(1 kHz). Trap-induced losses are shown not to limit measurements of other transitions.
\end{abstract}

\maketitle

\section{Introduction}

Calculable quantum field theory (QFT) corrections \cite{Horbatsch2016} and  precision laser spectroscopy \cite{hansch2006} have made hydrogen (H) an excellent low-energy system to probe fundamental physics \cite{Jones2020, Safronova2018, Karshenboim2010, Karshenboim2005, Brax2011, Burrage2018, Stadnik2018, Ghosh2020, delaunay2023}. Precise comparisons between H and anti-hydrogen ($\bar{\text{H}}$) also offer a stringent test of CPT symmetry \cite{Shore2005, Kostelecky2015, Ahmadi2017, Ahmadi2018}.

Fundamental physics tests in hydrogen rely on determining two parameters; the Rydberg constant $R_{\infty}$, which relates the theoretical energy scale to the SI system of units, and the ``size'' of the proton, characterized by its charge radius $r_p$, which cannot yet be accurately calculated from Quantum Chromodynamics \cite{Alexandrou2020}.
The 1S--2S transition frequency, which has been measured with a fractional uncertainty of $\sim 10^{-15}$\cite{Parthey2011, Matveev2013}, can be used to precisely determine one of these parameters, but must be complemented by measurements of other transitions or scattering-based measurements of $r_p$ \cite{Xiong2023}. It is well-known that the current dataset of relevant measurements is internally inconsistent \cite{Tiesinga2018, Mohr2014}, and inconsistent with measurements in muonic hydrogen \cite{Pohl2013}; a problem often called the ``proton charge radius puzzle" \cite{Gao2022}. Whilst this variation may be evidence of new physics (e.g. a hidden sector, light scalar boson \cite{Jones2020, Yost2022, delaunay2023}), the disagreement between two recent measurements of the 1S--3S transition \cite{Fleurbaey2018, Grinin2020} indicate that experimental systematics are at least partially responsible.

Currently, all precision H spectroscopy experiments rely on an atomic beam as the source \cite{Parthey2011, Matveev2013, Yost2022, Fleurbaey2018, Grinin2020, Bezginov2019, Beyer2017, Schwob1999, Beauvior1997, deBeauvoir2000}. Line shifts and broadening due to motional (Doppler) effects are a significant source of uncertainty, and careful velocity filtering \cite{Parthey2011} and/or lineshape analysis (see the supplementary information of \cite{Yost2022}) is required to extract precise measurements of the transition frequency. As a result, the overall uncertainty in measurements of the 1S--2S transition has not significantly advanced in 10 years, and motional effects may be a significant source of inconsistency within the H world dataset. 

In contrast, the precision spectroscopy of heavier neutral atoms has been revolutionised by the use of ultra-cold \textit{trapped} atoms in optical lattice clocks (OLCs) \cite{Derevianko2011, Takamoto2005, Le2006, Ludlow2006, Barber2006}. Here atoms are confined in a ``magic wavelength'' optical lattice such that trap-induced lineshifts cancel, and motional effects are eliminated by operating in the tight confinement (Lamb-Dicke) regime \cite{Derevianko2011}. OLCs have now reached a precision of $\sim 10^{-18}$, which surpasses the current definition of the SI second \cite{Bothwell2019}. An OLC operating on the dipole forbidden 1S--2S transition has been proposed as a route to improved measurement precision in $\bar{\text{H}}$ \cite{Crivelli2020}. The same clock operating in H would offer a new precision measurement of the 1S--2S transition with a different set of systematics. 

A magic wavelength for the 1S--2S transition is well established \cite{Adhikari2016, Adhikari2022} and the heating rate from elastic atom-photon scattering is known to be small \cite{Crivelli2020}. Quenching of the 2S meta-stable state by D.C. electric fields is also well known \cite{Holt1972}. Similar effects in off-resonant A.C. fields have the potential to limit coherence times and depopulate the excited state \cite{Dorscher2018}.
Differential cross sections have been calculated for 2S--1S inelastic scattering across a range of wavelengths \cite{Zernik1964, Gontier1971, Klarsfeld1972, Bachau2017}. The same is true for 2S--3S/D scattering very close to the known magic wavelength \cite{Heno1980}. Two-photon ionisation from the 2S state of hydrogen is similarly well understood \cite{zernik1964two, rapoport1969, Gontier1971, khristenko1976, Heno1980, kassaee1988, karule2003}. Despite this detailed theoretical attention, these effects are missing from recent discussions of trapped H measurement.

In this paper we give a comprehensive treatment of the effect of trap-induced elastic and inelastic light scattering and multi-photon ionisation on measurements involving the 2S state of H. We pay special attention to rates at four 1S--2S magic wavelengths --- one known and three newly reported.
We find excellent agreement with existing calculations of the 2S scattering rate and extend these works to consider all relevant final states at lower wavelengths.
At trap intensities relevant to the operation of an OLC, we find that two-photon ionisation provides the dominant loss channel. The result is a considerable reduction in the 2S lifetime leading to substantial broadening of the 1S-2S transition.
We calculate the minimal broadening compatible with the Lamb-Dickie regime and compare it to the best atomic beam 1S--2S measurement.
We also discuss the impact of trap induced losses from the 2S state on measurements of 2S--Rydberg transitions. We note that significant experimental challenges for cooling and trapping H exist \cite{VazquezCarson2022, Baker2021, Lane2015, Gabrielse2018, Burkley2018}; we do not consider them further here.

All mathematical expressions are given in Hartree atomic units(a.u., see, e.g., \cite{Drake2006}).
 
\section{Numerical calculations} 
\label{sec:numerical}

Central to any OLC is the optical lattice; a one dimensional lattice is described by the potential \cite{Grimm2000}:

\begin{align}
U(x) = U_0\cos^2{kx},
\end{align}
with
\begin{align}
    U_0 = -2\pi\alpha_\text{FS}\alpha_a(\omega)I_0,
\end{align}
where
$k$ and $\omega$ are the wave number and frequency of the lattice light respectively, $I_0$ is the peak lattice intensity, $\alpha_\text{FS}$ is the fine structure constant, and $\alpha_a$ is the polarisability of atomic state $a$. Throughout the text, we assume that the lattice light is linearly polarised in the $\hat{\textbf{z}}$ direction and that this matches the atomic axis of quantisation. It is often useful to characterise the depth of the lattice $|U_0|$ in terms of the single lattice photon recoil energy, $E_\text{rec} = \omega^2\alpha_\text{FS}^2/2m_\text{H}$,\footnote{This definition (in atomic units) is equivalent to\\ $E_\text{rec} = \hbar^2k^2/2m_\text{H}$.} giving a dimensionless lattice depth $D:= |U_0|/E_\text{rec}$.

\begin{figure}[]
    \centering
    \includegraphics[width=0.5\textwidth]{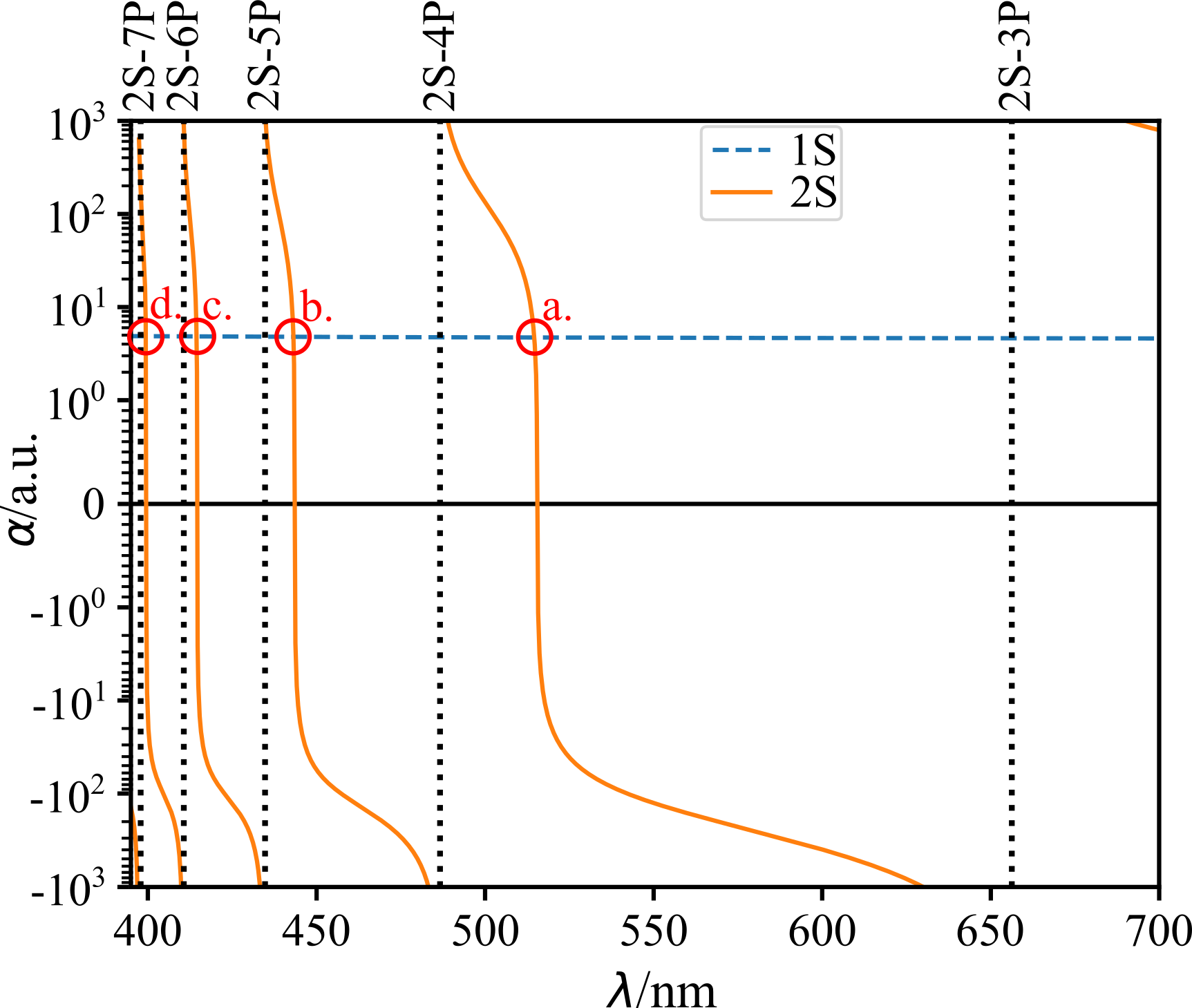}
    \caption{Polarisability of the 1S (blue dashed line) and 2S states (solid orange line) versus wavelength. A symmetric log scale is used on the y-axis. The four magic wavelengths listed in table 1 are labelled as a--d. Vertical dotted lines indicate 2S-nP atomic resonances.}
    \label{fig:magicwav}
\end{figure}

The polarisability of state  $a$ with zero orbital angular momentum, $l=0$ ($nS$ state) at frequency $\omega$ is given by \cite{Haas2006, LeKien2013},
\begin{align}
\label{eqn:pol}
\alpha_a(\omega) = \frac{1}{3}\sum_k\left(\frac{|r_{ka}|^2}{\omega_{k a} - \omega} + \frac{|r_{k a}|^2}{\omega_{ka} + \omega}\right).
\end{align}
The sum is across all states $k$ that are dipole coupled to $a$. Here ${r}_{ka}$ are the radial dipole matrix elements as defined in Appendix A (equation (\ref{eqn:raddef})) and $\omega_{ka} = \omega_k - \omega_a$ is the difference in energy between the states $k$ and $a$.  We calculate the atomic polarisability computationally with an implicit summation method (see Appendix B) in the non-relativistic theory of the hydrogen atom including the reduced mass. The leading order corrections to these values come from relativistic and field configuration terms and are of the order of $\sim \alpha_{FS}^2$ \cite{Adhikari2016, Adhikari2022}. Therefore values reported in this paper are quoted up to four significant figures. The polarisability of the 1S and 2S states is plotted for a range of optical wavelengths in figure \ref{fig:magicwav}.

Magic wavelengths, where the  polarisability of both states is equal, are critical for an OLC as they eliminate the differential light shift on the clock transition. To find these wavelengths, we numerically solve $|\alpha_\text{2S}(\omega) - \alpha_\text{1S}(\omega)| = 0$ in the region around each intersection labelled a--d in figure \ref{fig:magicwav}. We find four magic wavelengths in the considered range\footnote{For comparison it is useful to give these to more significant figures: $514.6464$, $443.212$, $414.483$, and $399.451$\,nm. The first agrees very well with the literature \cite{Adhikari2016, Adhikari2022}.} at $514.6$, $443.2$, $414.5$, and $399.5$\,nm. The polarisability at the magic wavelengths are reported in table \ref{tab:magicwav}.

\begin{table}[h!]
    \centering
    \caption{Key parameters for  1S-2S spectroscopy in magic wavelength traps:  polarisability $\alpha$ (in atomic units) of both states and the intensity that produces a 1$E_\text{rec}$ deep lattice. Values for the Rayleigh scattering rate $R_\text{el}$, the Raman scattering rate $R_\text{in}$ and the two-photon ionisation rate $R_\text{ion}$ are stated at a depth of $1\ E_\text{rec}$.}
    \setlength{\tabcolsep}{3pt}
    \begin{tabular}{c|c  c  c  c  c}
        \hline\hline \\ [-0.15cm]
       $\lambda$/nm & $\alpha$ & $I/\text{MWcm}^{-2}$ & $R_\text{el}/\text{s}^{-1}$  & $R_\text{in}/\text{s}^{-1}$ & $R_\text{ion}/\text{s}^{-1}$\\ [0.2cm]
       \hline \\ [-0.15cm]
        514.6 & 4.730 & 3.372 & $7.986\times10^{-3}$ & $61.49$ & $32.19$ \\ [0.1cm]
        
        443.2 & 4.813 & 4.460 & $17.16\times10^{-3}$ & $69.41$ & $30.29$ \\ [0.1cm]
        
        414.5 & 4.863 & 5.056 & $24.23\times10^{-3}$ & $73.90$ & $29.04$ \\ [0.1cm]
        
        399.5 & 4.890 & 5.413 & $29.31\times10^{-3}$ & $76.78$ & $28.33$ \\ [0.2cm]
        \hline
    \end{tabular}
    \label{tab:magicwav}
\end{table}

According to the definition of the magic wavelength, the polarisability is ultimately limited by that of the 1S state. Figure \ref{fig:magicwav} and table \ref{tab:magicwav} show that this polarisability is very small, approximately $4.8$\, a.u. at each magic wavelength compared to $\sim 280$\,a.u. in Sr at the 813\,nm magic wavelength \cite{Dorscher2018}, due to the absence of 1S resonances at wavelengths shorter than 121.6\,nm (the Lyman alpha line). In addition, the low mass of H leads to large recoil velocities at optical frequencies and substantial laser power is required to trap in the 1S state.
Optical lattice clocks usually operate in deep optical traps, often around $100E_\text{rec}$. Such a deep lattice for a hydrogen OLC requires intensities of 100's\,$\text{MW}/\text{cm}^2$ (see table \ref{tab:magicwav}), $O(10^4)$ times larger than a comparable lattice for Sr. Nevertheless,  such high intensities are achievable with current laser technology, particularly for 514.6\,nm where significant power is available from frequency-doubled 1029.2\,nm radiation \cite{Crivelli2020, Alnis2008}.

A major concern in optical trapping is off-resonant atom-photon scattering. It is useful to separate the various scattering processes by the final internal state of the atom. Firstly we consider elastic or Rayleigh scattering, which does not change the internal state. The key effects of elastic scattering is to change the vibrational state of trapped atoms, resulting in heating \cite{Grimm2000}. The rate of elastic photon scattering is closely related to the polarisability and can be written as (see Appendix \ref{app:scattering}),
\begin{align}
\label{eqn:rayleigh}
R_\text{el} = \omega^3\alpha_\text{FS}^4\frac{8\pi}{3}|\alpha_a(\omega)|^2I.
\end{align}
By the definition, this rate is the same for both the 1S and 2S states in a magic wavelength trap. Therefore, we present a single value for the elastic scattering rate in table \ref{tab:magicwav}.
These rates are very small, only approaching the spontaneous decay rate at around $1000 E_\text{rec}$ at $514.6$\,nm ($270 E_\text{rec}$ at $399.5$\,nm).

Next, we consider inelastic scattering to other bound internal states. Critically, this includes scattering directly to the ground state. These processes proceed via all intermediate states $k$ that are dipole coupled to both the initial state $a$ and some final state $b$. As explained in appendix \ref{app:scattering}, the scattering rate from an initial S state $a$ to final state $b$ can be expressed in the following form:
\begin{align}
\label{eqn:scattering_angle}
R_{ba} = \omega_s^3\alpha^4_\text{FS}\mathcal{A}_{ba}\left(\sum_k \frac{r_{b k}r_{ka}}{\omega_{ka} \mp \omega} + \frac{r_{b k}r_{ka}}{\omega_{b k} \pm \omega} \right)^2 I,
\end{align}
where $\omega_s = - \omega_{ba} \pm \omega$ is the angular frequency of the scattered photon and $\mathcal{A}_{ba}$ is an angular factor (see Appendix \ref{app:scattering}). Dipole selection rules restrict scattering to only S or D final states. Upper and lower signs relate to Raman scattering (RS) and singly stimulated two-photon emission (SSTPE) \cite{Braunlich1970} respectively. A breakdown of the total 2S inelastic scattering rate according to final gross state (summing over magnetic sub-levels) is given in table \ref{tab:Ramanrate} and indicates that direct scattering to the ground state is the dominant process. 

The rates of both elastic and inelastic scattering were calculated with the same implicit summation method as the polarisability (Appendix \ref{app:sturmian}). The elastic scattering rate is consistent with existing work, while results for scattering to 1S, 3S and 3D states at 514.6\,nm are in good agreement with previous calculations close to this value \cite{Heno1980}. The results for additional final states and other magic wavelengths are new, to the best of our knowledge. Leading order corrections from relativistic and field configuration terms are of the same order as for polarisability.

To obtain the total rate of inelastic atom-photon scattering for state $a$, we sum RS and SSTPE rates for all allowed final states,
\begin{align}
    R_\text{in} = \sum_{b\neq a}^{\omega_{ba} < \omega} R_{ba}.
\end{align}
The 2S inelastic scattering rates at the magic wavelengths are also presented in table \ref{tab:magicwav}. These rates are much larger than the elastic scattering rates at the same wavelengths. In particular, they exceed the spontaneous decay rate for depths as low as $O(0.1)E_\text{rec}$.

\begin{table*}
    \centering
    \caption{Rates for allowed inelastic scattering processes with nS and nD final states (listed according to final state) and total inelastic scattering rates for the 2S initial states. The rates are given per unit depth in units of s$^{-1}$ at a depth of 1$E_\text{rec}$ in the 2S state.}
    \setlength{\tabcolsep}{6pt}
    \begin{tabular}{c|c c c c c c c c c c}
        \hline\hline \\ [-0.15cm]
        $\lambda$/nm & 1S & 3S & 3D & 4S & 4D & 5S & 5D & 6S & 6D &  $R_\text{in}/D$  \\ [0.2cm]
        \hline \\ [-0.15cm]
        514.6 & $57.32$ & $3.843$ & $0.3329$ & - & - & - & - & - & - & $61.49$  \\ [0.1cm]
        443.2 & $62.85$ & $3.269$ & $0.1861$ & $2.295$ & $0.8110$ & - & - & - & - & $69.41$  \\ [0.1cm]
        414.5 & $65.75$ & $3.249$ & $0.1570$ & $1.892$ & $0.5481$ & $1.467$ & $0.8388$ & - & - & $73.90$  \\ [0.1cm]
        399.5 & $67.53$ & $3.272$ & $0.1453$ & $1.802$ & $0.4636$ & $1.207$ & $0.6109$ & $1.005$ & $0.7460$  & $76.78$ \\ [0.2cm]
        \hline
    \end{tabular}
    \label{tab:Ramanrate}
\end{table*}

\begin{figure}[h]
    \includegraphics[width=0.45\textwidth]{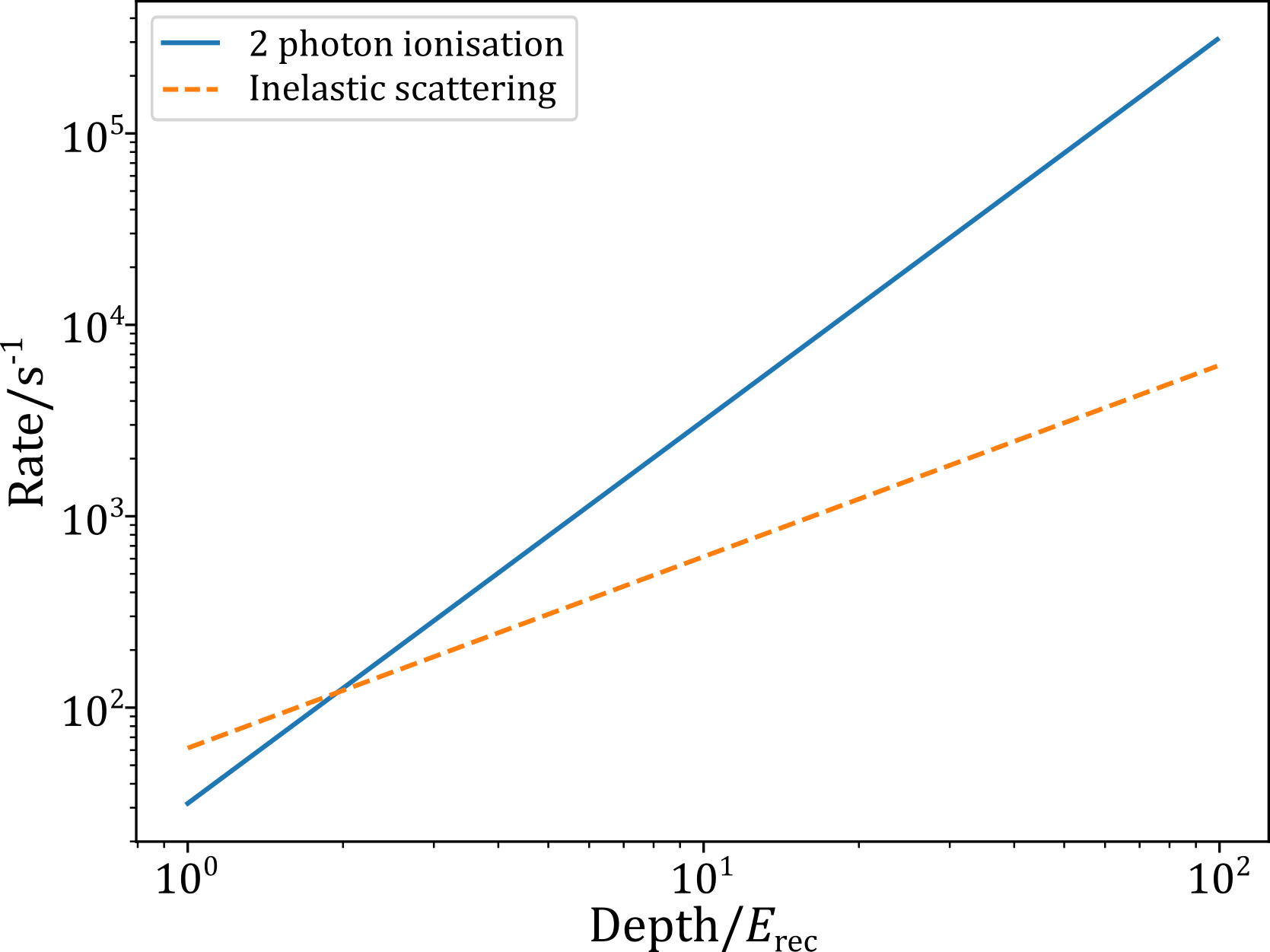}
    \caption{Comparison between the two-photon ionisation rate (solid blue lines) and the inelastic scattering rate (dashed orange lines) for varying trap depth at 514.6\,nm. At this scale, these plots look the same for all magic wavelengths. The intersection between the two lines does not change substantially either: it is at $2E_\text{rec}$ for 514.6\,nm and $3.2E_\text{rec}$ for 399.5\,nm.}
    \label{fig:variations}
\end{figure}

Lastly we consider inelastic scattering to continuum states, resulting in ionisation. Single photon ionisation from the 2S state is only possible at wavelengths below 365\,nm. Ionisation at the magic wavelengths thus involves absorption of at least two photons. 
We use the STRFLO program \cite{Potvliege1998} to calculate multi-photon ionisation rates.
The two-photon ionisation rates at the magic wavelengths are given in the final column of table \ref{tab:magicwav} for a specified depth of $1E_\text{rec}$. Inelastic scattering rates scale linearly with intensity, while these ionisation rates scale with intensity squared. Therefore, ionisation quickly dominates inelastic scattering as trap depth increases past O(1)$E_\text{rec}$, as shown in figure \ref{fig:variations}. The dominance of ionisation for relatively low trap depths is a result of the low polarisability of the 1S state, and the high trapping intensities that result.

Table \ref{tab:magicwav} and figure \ref{fig:variations} indicate that at magic wavelengths, the impact of trap induced ionisation is enormous. At a depth of $1\ E_\text{rec}$ the 2S state lifetime is reduced from 125\,ms to O(10)\,ms, while at $100E_\text{rec}$ (a common depth for an OLC) it becomes just O$(1)\unit{\micro\second}$. These results impose severe limitations on the coherence times that can be achieved in a H lattice clock.

\begin{figure*}
    \centering
    \includegraphics[width=0.95\textwidth]{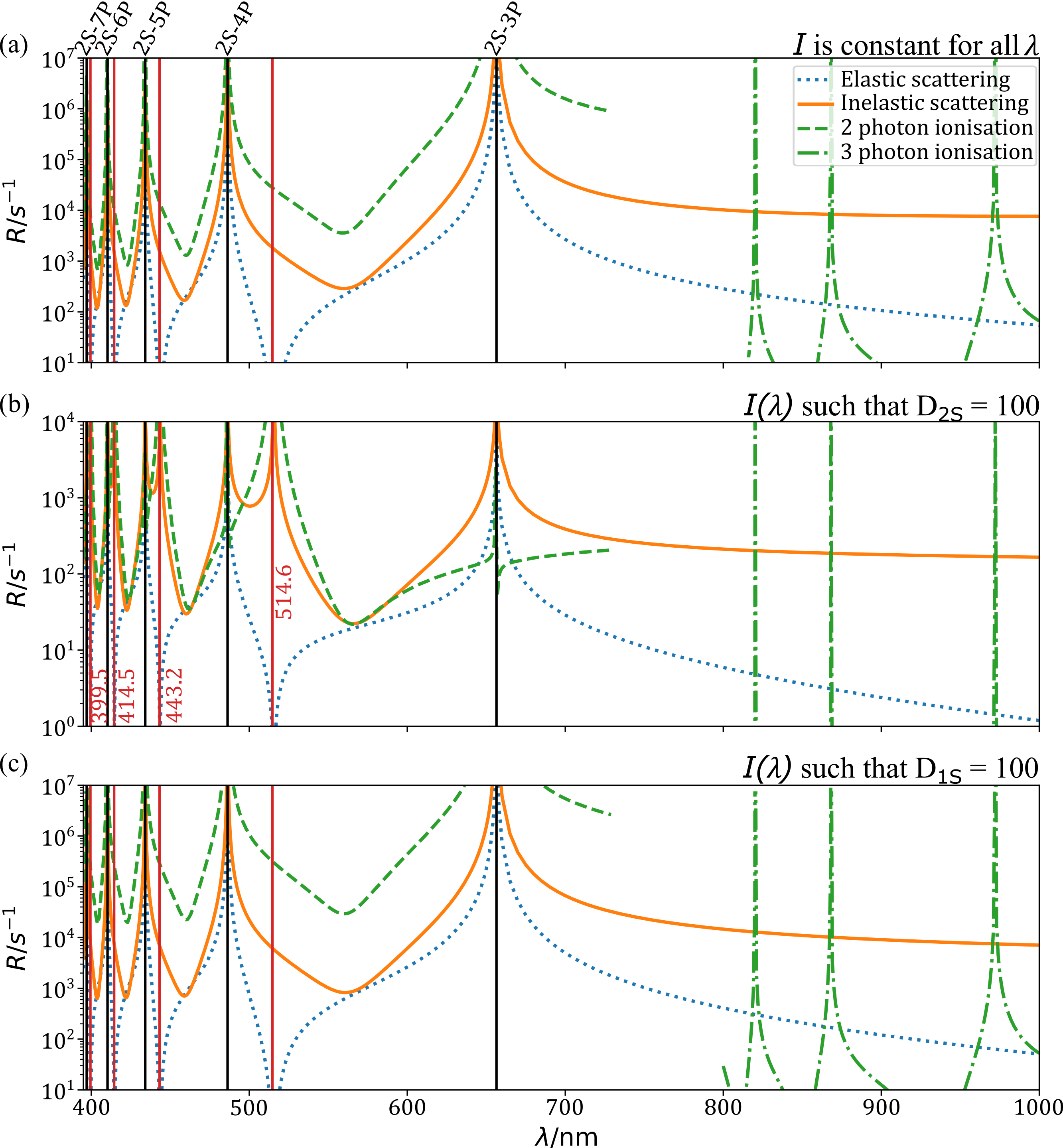}
    \caption{The variation of elastic scattering (dotted blue lines), inelastic scattering (solid orange lines), and two-photon ionisation (dashed green lines) rates from the 2S state across a range of wavelengths. The line showing two photon ionisation only extends to the threshold. Also plotted are the three-photon ionisation rates (dashdotted lines) for wavelengths 800-1000nm, which is the leading order ionisation process in this region. (a) Rates at constant intensity  $I = 100\  \text{MW}\text{cm}^{-2}$. (b) and (c) Rates at variable intensities which maintain a constant $D=100$ lattice for depths defined with reference to the 2S and 1S states respectively. Black vertical lines indicate 2S atomic resonances and red vertical lines indicate magic wavelengths.}
    \label{fig:comparisons}
\end{figure*}

The impact of operating at a magic wavelength is highlighted by plotting the scattering and ionisation rates as a function of wavelength, as in figure \ref{fig:comparisons}. In figure \ref{fig:comparisons}(a)  the rates are shown for a constant intensity of $100\,\text{MW\,cm}^{-2}$; corresponding to trap depths ranging from $30E_\text{rec}$ at 514.6\,nm to $18E_\text{rec}$ at 399.5\,nm. Here, peaks in the rates correspond to the  resonances of the Balmer series. Two-photon ionisation does not extend past the threshold at $729$\,nm. For longer wavelengths, ionisation proceeds via the absorption of at least three photons. The 3-photon ionisation rates vary as $I^3$ but are generally much smaller than the inelastic scattering rate at these intensities, except at narrow 2 photon resonances with intermediate bound states.

In contrast, figures \ref{fig:comparisons}(b) and (c) show the rates at a constant trap depth $D=100$. In (b) it is the 2S trap depth that is fixed, while in (c) it is the 1S trap depth. We note that  $D$ ignores the sign of the potential; in wavelength regions where the polarisability is negative (blue detuned) atoms are trapped at intensity minima and the actual loss rate observed in experiment may be lower. Compared to figure \ref{fig:comparisons}(a), figure \ref{fig:comparisons}(b), shows an extra series of divergences in the inelastic scattering and 2-photon ionisation rates. These originate from zero-crossings of the 2S polarisability. Here the intensity required to produce a trap at a given depth, and hence the rates, diverge. The magic wavelengths sit very close to these zero-crossings, suppressing elastic scattering, but enhancing inelastic scattering and ionisation at a given trap depth. The 1S polarisability is essentially flat in this region and does not cross zero, so the shape of figure \ref{fig:comparisons}(c) is very similar to part (a). Compared to part (b) the rates are generally much larger, except at the magic wavelength. Again, this is a result of the small, almost constant polarisability of the 1S state.

\section{Discussion}
\label{sec:discussion}
First, we consider a 1S--2S hydrogen OLC, operating in a magic wavelength lattice. As seen in the previous section, the lifetime $\tau$  of the 2S state is severely reduced by two-photon ionisation at these wavelengths. This broadens the minimal linewidth of the 1S--2S transition $\Gamma_\text{FWHM} = 1/(2\pi \tau)$. This effect is considerable in the deep lattices that are usual for OLCs. For example, in a $100E_\text{rec}$ deep magic wavelength lattice, the 2S state lifetime is reduced to O(1)$\unit{\micro\second}$, broadening the line from a natural width of $1.27$\ Hz to an ionisation dominated $\sim 50$\ kHz. 

It is instructive to make a comparison to measurements of 1S--2S in cold atomic beams of H \cite{Parthey2011, Matveev2013}. The systematic uncertainty in these measurements is dominated by velocity-dependent effects such as the second-order Doppler shift. To reduce these effects, the measurement must selectively address atoms that sit within a very narrow range in the slow tail of the velocity distribution. This results in a momentum-dependent asymmetric lineshape with a measured width of $\sim2$\,kHz \cite{Parthey2011}.

An important motivation for using lattice-trapped atoms is the potential to eliminate the velocity dependence of the lineshape. To achieve this we must operate in the limit of resolved sidebands, where the trap frequency (given in SI units),
\begin{align}
    \nu_\text{T} = \sqrt{D}\frac{h}{m_\text{H}\lambda^2},
\end{align}
is much larger than the linewidth, $\nu_\text{T} \gg \Gamma_\text{FWHM}$. In this limit, the effects of changes in vibrational state are separated into distinct sideband signals. This leaves a central carrier line that is insenstive to motional effects and symmetric in the non-relativistic limit. The low mass of H and relatively short magic wavelengths result in large trap frequencies, 
\begin{align}
    \nu_\text{T} = \sqrt{D} \times \begin{cases} 
    1.50\, \text{MHz  at  } 514.6 \,\text{nm} \\
    2.02\, \text{MHz  at  } 443.2 \,\text{nm} \\
    2.31\, \text{MHz  at  } 414.4 \,\text{nm} \\
    2.48\, \text{MHz  at  } 399.4 \,\text{nm}. 
    \end{cases}
\end{align}
This compares favourably with the linwidth and makes it simple to achieve well-resolved sidebands. In fact, the broadened linewidth only becomes comparable to the trap frequencies for depths greater than O(1000)$E_\text{rec}$ or lower than $O(10^{-11})E_\text{rec}$. 

It is also desirable to operate a clock in the Lamb-Dicke regime, where the separation of vibrational states is much larger than the recoil energy of a 243\,nm probe photon, $h\nu_\text{T} \gg E_\text{probe}$. This suppresses transitions to other vibrational states and reduces the relative size of the side band signals compared to the carrier.
The 1S--2S transition in hydrogen is a 2-photon transition and can be driven in a Doppler-free manner at 243\,nm \cite{hansch1975}. 
Whilst this suppresses changes in the atom's vibrational state to first order, higher-order effects due to wavefront curvature remain. These momentum-changing effects appear in the second-order sidebands (see Appendix \ref{app:LDregime}). To make a quantitative statement, we demand that the probability of remaining in the ground vibrational state $P_{n=0\rightarrow n=0} > 0.9$ (equivalent to $\eta^2 < 0.381$), and find the results listed in table \ref{tab:LDdepths}. 

However, the carrier remains subject to a small shift due to second-order Doppler effect, which arises from time dilation between the lab frame and the moving atom frame \cite{biraben1991}. Following the calculation of \cite{martinez2022}, and assuming a trap frequency of $8.79$\,MHz (see table \ref{tab:LDdepths}) we obtain a fractional frequency shift of $-9.68\times10^{-18}$ in the vibrational ground state --- an absolute shift of $-23.9$\,mHz to the 1S--2S transition frequency (see Appendix \ref{app:LDregime}). Similar calculations for thermal vibrational states result in a fractional shift of $O(10^{-17})$, which is well below the current uncertainty in the 1S--2S measurement \cite{Parthey2011}. 

These results show that effective control of velocity dependent systematics can be achieved in much shallower lattices than is usual in other species. This allows for linewidths that are comparable to, or better than, those of beam measurements, but with the advantage that velocity-dependent systematics are reduced for the carrier signal. The narrower linewidths attainable at shorter magic wavelengths result from the tighter confinement in these lattices.

\begin{table}[h!]
\centering
\label{tab:LDdepths}
\caption{The trap depth and 1S--2S linewidth at the intensity where the Lamb-Dicke constraint is met for each magic wavelength. These depths all relate to trap frequencies of 8.79\,MHz and a Lamb-Dicke parameter of $\eta^2 < 0.381$.}
    \setlength{\tabcolsep}{3pt}
    \begin{tabular}{c | c c}
     \hline\hline \\ [-0.15cm]
        $\lambda$\,/nm & $D$ & $\Gamma_\text{FWHM}$\,/kHz\\[0.2cm]
        \hline \\ [-0.15cm]
        514.6 & 34.5 & 6.43 \\[0.1cm]
        443.2 & 19.0 & 1.95 \\[0.1cm]
        414.5 & 14.5 & 1.14 \\[0.1cm]
        399.5 & 12.5 & 0.857 \\ [0.2cm]
        \hline
    \end{tabular}
\end{table}

It should be noted that the above discussion treats each lattice site as a harmonic potential. This is a good approximation at the bottom of the lattice site but less so for higher-lying vibrational states. Additionally, we have assumed that the driving field aligns with the trapping field and considered only motion in this one dimension.

Compared to atomic beams, trapped atom experiments usually sacrifice statistical power in favour of much narrow linewidths. For a hydrogen OLC however, the best achievable linewidths will remain comparable to those available in beams --- which still hold a significant statistical advantage.
In Sr optical lattice clocks, statistical power is improved by reading out fluorescence from a fast-cycling transition out of the ground state \cite{Derevianko2011}. Unfortunately, due to the lack of laser power at Lyman series wavelengths \cite{Gabrielse2018}, such a scheme is impractical in H.
At the magic wavelengths, trap-induced ionisation will provide a continuous readout of the 2S state population, with the drawback that it is destructive and so the trap must be replenished. 
Currently, cooling and loading atomic H into an optical trap is an open problem, but it is clear that optimising the experimental duty cycle is crucial in minimising the statistical uncertainty that can be achieved. 

An alternative to working at a magic wavelength is to move to longer wavelengths where two-photon ionisation is suppressed\footnote{The Lamb shift implies that a very long magic wavelength should exist between the 2S-3P and the 2S-$\text{2P}^{1/2}/\text{2P}^{3/2}$ resonances, but this cannot be obtained by the non-relativistic theory alone.}. However, using a non-magic trap introduces intensity-dependent systematics to the measurement exacerbated by the large differential light shift across the transition (e.g. the 2S polarisability is around 40 times larger than the 1S polarisability at 1064\,nm). Also, the need to trap atoms in the 1S state as well as the 2S state means broadening from inelastic scattering is still significant (compare figure \ref{fig:comparisons}(b) to (c)). In addition, the lower trap frequencies available at long wavelengths compared to the magic wavelengths mean that even deeper traps are required to enter the Lamb-Dicke regime.

We now briefly consider the implications of these results for other transitions in atomic hydrogen. The 2S state provides a suitable spectroscopic ground state with accessible transitions to many states of higher principal quantum number $n$ (e.g. \cite{Yost2022, Bezginov2019, Beyer2017, deBeauvoir2000}). These states decay much more quickly than the 2S state, so their lifetime dominates the transition linewidth. This is particularly true for low-$n$ states with lifetimes of O(10)\,ns \cite{jitrik2004transition}. High-$n$ states are longer lived --- at $n\approx30$, states with low $l$ have lifetimes O(10)$\,\unit{\micro\second}$, and states with high $l$ can have lifetimes\footnote{These lifetimes are often discussed in the context of Rydberg-Stark states: see \cite{seiler2016} for a discussion of these lifetimes or \cite{Vliegen2007} for the lifetime of the $n=35$, $k=30$, $m=0$ state.} of 100's $\unit{\micro\second}$. Since the trap depth for the 1S state is no longer essential, one can trap with much lower intensities away from the 1S--2S magic wavelengths. This significantly reduces the rates of two-photon ionisation and inelastic scattering for traps at a given depth (see figure \ref{fig:comparisons}(b)). For example, even a very deep trap of $D=100$ (for the 2S state) only quenches the 2S lifetime to $\sim 6$\,ms at 1000\,nm.

In particular transitions to high-lying Rydberg states \cite{Jones2020} have much to gain from trapped atom measurement. These high-$n$ states exhibit strong dipole-dipole interactions which may introduce significant uncertainty into beam or vapour measurements. The well-defined inter-atomic spacing provided by an optical lattice or a tweezer array would enable control and characterisation of these interactions. Field-free spectroscopy can be performed by briefly turning off the trap potential \cite{Sassmannhausen2013}. In this case, the trap-induced broadening of the 2S state does not limit the measured linewidth. Instead, the minimal achievable linewidth is primarily limited by the natural lifetime of the Rydberg state.  The reduced 2S lifetime only serves to limit the accumulation of 2S population.

Finally, we mention the implications for measurements in deuterium (D) and $\bar{\text{H}}$. First, the non-relativistic theory of H and D is identical up to a difference in the reduced mass, $\mu$, of $(\mu_\text{D} - \mu_\text{H})/\mu_\text{H} = 2.702\times10^{-4}$. Thus the conclusions drawn for H are valid for D, with relative differences in exact values O($10^{-4}$). 
The structure of H and $\bar{\text{H}}$ are identical in non-relativistic quantum mechanics. As such, the results presented in this paper also hold for $\bar{\text{H}}$ with the same level of precision.  Atomic beams of $\bar{\text{H}}$ are highly impractical, so the 1S--2S transition is instead measured in a dilute gas confined in a deep magnetic trap \cite{Ahmadi2018, Ahmadi2017}. 
Whilst recent measurements in this system have seen linewidths of 5~kHz\cite{Ahmadi2018}, the limited number of anti-atoms available limits the total fractional uncertainty to $2\times10^{-10}$~\cite{Ahmadi2017}.

\section{Conclusions}
\label{sec:summary}
We have calculated the polarisability of the 1S and 2S states in atomic hydrogen and identified new magic wavelengths in the range 395-1000\,nm (figure \ref{fig:magicwav}). We have also calculated the atom-photon scattering and two-photon ionisation rates out of the 2S state across in this wavelength range (figure \ref{fig:comparisons}), paying particular attention to rates at the magic wavelengths (tables \ref{tab:magicwav} and \ref{tab:Ramanrate}).

Two-photon ionisation significantly broadens the 1S--2S transition linewidth in deep magic wavelength traps. This broadening is a consequence of the low polarisability of the 1S state and the resultant high trapping intensities. 
However, the low mass of atomic hydrogen allows for high trap frequencies. This makes it possible to enter the resolved sideband and Lamb-Dicke regimes in relatively shallow traps compared to heavier atoms like Sr (see table \ref{tab:LDdepths}), opening a route to spectroscopy free from momentum-dependent systematics. 
In these shallower traps, the effect of ionisation is no longer catastrophic, and linewidths of $\sim 1$ kHz are achievable, especially at the shorter magic wavelengths.

It is unlikely that a 1S--2S lattice clock will be competitive as an absolute frequency reference when compared to Sr lattice clocks or modern ion clocks. It could offer a measurement of the 1S--2S transition in H, D and even $\bar{\text{H}}$ with a narrow line, free from velocity-dependent systematics. The precision that can be achieved will depend critically on the atom number and duty cycle. Comparisons between clock measurements in H, D, and $\bar{\text{H}}$ would set powerful constraints on possible physics beyond the Standard Model.

There is no need to produce deep traps of the 1S state for spectroscopy out of 2S. This allows deep 2S traps at wavelengths far from the magic wavelengths and at lower trapping intensities, and therefore with reduced ionisation and inelastic scattering rates. Given that the lifetimes of other hydrogen states are generally shorter than the 2S, the 2S state lifetime does not limit the linewidth in any reasonable trap. Trapped atom measurements are particularly promising for measurements to high-lying Rydberg states. Here, the well-defined inter-atomic spacing allows for control of systematics related to strong dipole-dipole interactions. 

\begin{acknowledgments}
J. P. Scott is supported by a Stubbs Scholarship, and we gratefully thank Rodney and Francis Stubbs for their support. We thank Dylan Yost and Thomas Udem for useful comments and advice.
\end{acknowledgments}

\appendix
\section{Scattering rates}
\label{app:scattering}

The differential cross section for atom-photon scattering is given by the Kramers-Heisenberg formula \cite{Loudon2010}:
\begin{align}
\begin{gathered}
\frac{\text{d}\sigma}{\text{d}\Omega} = \sum_{b}^{\omega_{ba} < \omega}\omega(\pm\omega - \omega_{ba})^3\alpha_\text{FS}^4 \\ \times \left|\sum_k\left(\frac{\boldsymbol{\epsilon}^*_s\cdot {\bf r}_{b k}\boldsymbol{\epsilon}\cdot {\bf r}_{ka} }{\omega_{ka} \mp \omega} + \frac{\boldsymbol{\epsilon}\cdot {\bf r}_{b k}\boldsymbol{\epsilon}^*_s\cdot {\bf r}_{ka} }{\omega_{kb} \pm \omega}\right)\right|^2,
\end{gathered}
\end{align}
where $\boldsymbol{r}_{ka}$ is the dipole matrix element $\bra{k}\boldsymbol{r}\ket{a}$ and $\omega_{ka} = \omega_{k} - \omega_{a}$ is the energy difference between atomic states $k$ and $a$.
This scattering involves two photons; $\boldsymbol{\epsilon}$ is the polarisation vector of the trap photon --- taken to be linear --- and $\boldsymbol{\epsilon}_s$ is that of the scattered photon.
The polarisation of the scattered photon can be in any direction normal to its direction of emission, given by the co-latitude angle $\vartheta$ and the azimuthal angle $\varphi$. Therefore, it is necessary to sum across two orthogonal vectors which span the space of polarisation states for the scattered photon: $\boldsymbol{\epsilon}_1$ and $\boldsymbol{\epsilon}_2$. The polar angles of these vectors are denoted as $\vartheta_1$ and $\varphi_1$ for $\boldsymbol{\epsilon}_1$, and $\vartheta_2$ and $\varphi_2$ for $\boldsymbol{\epsilon}_2$; we choose:
\begin{align}
    \vartheta_1 = \vartheta - \frac{\pi}{2} &\text{ , } \varphi_1 = \varphi,\\
    \vartheta_2 = \frac{\pi}{2} &\text{ , } \varphi_2 = \varphi - \frac{\pi}{2}.
    \label{eqn:pol_angles}
\end{align}

The total atom-photon scattering rate can then be found,
\begin{align}
\label{eqn:ratebreakdown}
R = \int \text{d}\Omega \sum_{\boldsymbol{\epsilon}_s}\frac{\text{d}\sigma}{\text{d}\Omega}\frac{I}{\omega} = \sum_b^{\omega_{ba} < \omega} R_{ba},
\end{align}
where $R_{ba}$ is the rate of scattering which drives the two-photon allowed atomic transition $a \rightarrow b$,
\begin{align}
\begin{split}
    R_{ba} = \omega_s^3\alpha^4_\text{FS}\int \text{d}\Omega \sum_{s=1,2} \left|\sum_k \frac{(\boldsymbol{\epsilon}_s^*\cdot\boldsymbol{r}_{b k})(\boldsymbol{\epsilon}\cdot\boldsymbol{r}_{ka})}{\omega_{ka} \mp \omega} \right. \\ \left. + \frac{(\boldsymbol{\epsilon}\cdot\boldsymbol{r}_{b k})(\boldsymbol{\epsilon}_s^*\cdot\boldsymbol{r}_{ka})}{\omega_{kb} \pm \omega} \right|^2 I.
\end{split}
\label{eqn:appscatter}
\end{align}
Scattering may proceed via the absorption of a trap photon and emission of a scattered photon of frequency $\omega_s = -\omega_{ba} + \omega$ (Raman scattering --- RS), or via the emission of both a lattice frequency photon and a photon of frequency $\omega_s = -\omega_{ba} - \omega$ (Singly stimulated two photon emission --- SSTPE). The upper signs relate to RS, and the lower signs to SSTPE --- which is only possible when $b$ is lower in energy than $a$.

We will now pay particular attention to sums over intermediate states, this sum can be split into two distinct sums:
\begin{align}
    \sum_k \frac{(\boldsymbol{\epsilon_s^*}\cdot\boldsymbol{r}_{b k})(\boldsymbol{\epsilon}\cdot\boldsymbol{r}_{ka})}{\omega_{ka} \mp \omega} + \sum_k \frac{(\boldsymbol{\epsilon}\cdot\boldsymbol{r}_{b k})(\boldsymbol{\epsilon_s^*}\cdot\boldsymbol{r}_{ka})}{\omega_{kb} \pm \omega},
\end{align}
which we will refer to as sums (i) and (ii).

We begin by defining a basis of spherical unit vectors:
\begin{align*}
\hat{\epsilon}_{-1} = \frac{\hat{x} - i\hat{y}}{\sqrt{2}}, \hspace{1cm} \hat{\epsilon}_0 = \hat{z}, \hspace{1cm} \hat{\epsilon}_{+1} = -\frac{\hat{x} + i\hat{y}}{\sqrt{2}}.
\end{align*} The dipole operator may be written in this basis as,
\begin{align}
\hat{\epsilon}_q\cdot\boldsymbol{r} = \sqrt{\frac{4\pi}{3}}rY_{1q}(\theta, \phi),
\end{align}
where $\theta$ and $\phi$ are the polar angles of $\boldsymbol{r}$. This operator is clearly separable into radial and angular parts, the angular matrix elements are given,
\begin{align}
\begin{split}
 &A_{l'm'q, lm} = \\ 
 &\sqrt{\frac{4\pi}{3}}\int^\pi_0\sin\theta d\theta \int^{2\pi}_0d\phi Y^*_{l'm'}(\theta, \phi)Y_{1q}(\theta, \phi)Y_{lm}(\theta, \phi) \\
 &= (-1)^{-(m + q)}\sqrt{(2l+1)(2l'+1)}\times \\&\hspace{2cm}\begin{pmatrix} l' & 1 & l \\ 0 & 0 & 0 \end{pmatrix}\begin{pmatrix} l' & 1 & l \\ -(m+q) & q & m \end{pmatrix}\delta_{m', m+q}.
 \end{split}
 \label{eqn:angular}
\end{align}
We restrict consideration to trap photons that are linearly polarised along the $\hat{\boldsymbol{\epsilon}}_0$ (or $\hat{\boldsymbol{z}}$) direction, so:
\begin{align}
    \boldsymbol{\epsilon}\cdot\boldsymbol{r}_{bk} =  A_{l_bm_b0,l_km_k}r_{bk},
\end{align}
and the same for $\boldsymbol{\epsilon}\cdot\boldsymbol{r}_{ka}$.
Here, $r_{bk}$ represent the radial matrix elements:
\begin{align}
    r_{bk} = \int^\infty_0r^2dr R_b^*(r) r R_k(r).
    \label{eqn:raddef}
\end{align}
where $R_b(r)$ is the radial wave function of atomic state $b$.

In general, the scattered photon is not restricted to linear polarisation. The generic polarisation vector can be written in the same basis, 
\begin{align}
\boldsymbol{\epsilon} = \sum_{q\in\{-1, 0, 1\}}\sqrt{\frac{4\pi}{3}}Y_{1q}(\vartheta, \varphi)(-1)^{q}\hat{\epsilon}_{-q}.
\end{align}
Then, 
\begin{align}
\boldsymbol{\epsilon_s}\cdot\boldsymbol{r}_{bk} = r_{bk}\sum_{q\in\{0, \pm1\}}(-1)^q\sqrt{\frac{4\pi}{3}}Y_{1q}(\vartheta_s, \varphi_s)A_{l_bm_b-q, l_km_k},
\end{align}
and the same for $\boldsymbol{\epsilon}_s\cdot\boldsymbol{r}_{ka}$.

We consider only the case where $a$ is an S state. This restricts $k$ to P states only and so the angular term is the same for all intermediate states with a given final state.
The angular terms of the two sums differ only by the terms $A_{l_bm_b-q,1m_k}A_{1m_k0,00}$ for (i) and $A_{l_bm_b0,1m_k}A_{1m_k-q, 00}$ for (ii). We are interested in possible equality of these two terms. 
$A_{l_bm_b-q,1m_k}A_{1m_k0,00}$ and $A_{l_bm_b0,1m_k}A_{1m_k-q, 00}$ can be written explicitly using equation (\ref{eqn:angular}). Applying the delta functions and ignoring common terms leaves the possible equality:
\begin{align}
    \begin{pmatrix} l_b & 1 & 1 \\ q & -q & 0\end{pmatrix}\begin{pmatrix} 1 & 1 & 0 \\ 0 & 0 & 0\end{pmatrix} \overset{!}{=} (-1)^q\begin{pmatrix} l_b & 1 & 1 \\ q & 0 & -q\end{pmatrix}\begin{pmatrix} 1 & 1 & 0 \\ q & -q & 0\end{pmatrix}.
\end{align}
For $l_b$ an even number (assured for an initial S state), symmetries of the Wigner 3-j symbol mean that the first symbol on the LHS is equal to the first symbol and sign term on the RHS. One can then check the remaining symbol for each allowed value of $q \in {0, \pm1}$ and see that the equality holds. One can also check that $A_{l_bm_b1, 1m_k}=A_{l_bm_b-1,1m_k}$.
Therefore, all terms in both sums share a common angular term, and equation (\ref{eqn:appscatter}) can be recast as equation (\ref{eqn:scattering_angle}):
\begin{align}
    R_{ba} = \omega_s^3\alpha_\text{FS}^4\mathcal{A}_{ba}\left(\sum_k\frac{{r}_{b k}{r}_{ka} }{\omega_{ka}\mp \omega} + \frac{{r}_{bk}{r}_{ka}}{\omega_{kb}\pm\omega}\right)^2I,
\end{align}
where the angular term is given by:
\begin{widetext}
\begin{align}
    \mathcal{A}_{ba} =\int d\Omega \left[ \sum_{s = 1,2}\sum_{q \in \{0, \pm 1\}}\left|\sqrt{\frac{4\pi}{3}}(-1)^{q}Y_{1q}^*(\vartheta_s, \varphi_s)A_{l_bm_b-q,1m_k}A_{1m_k,00}\right|^2\right].
\end{align}

Terms in the integrand are vanishing in all cases except when $m_k = 0$, and $m_b = m_k - q = -q$ due to the delta functions present in \ref{eqn:angular}.
As such, for a final state $b$ with well defined $m_b$, only a single vale of $q$ contributes to the above integral. Further, scattering is allowed for final states with $m_b \in \{0, \pm 1\}$ only.

It is often desirable to consider the total scattering rate to a given gross state with specified $n$ and $l$, as in table \ref{tab:Ramanrate}. In this case it is necessary to sum over the magnetic sub-levels of $b$ (in principle one must also sum over $m_a$, but we have restricted $m_a$ to 0 only). Writing out this sum explicitly and apply the definitions of $\vartheta_s$ and $\varphi_s$ in \ref{eqn:pol_angles}, we obtain,
\begin{align}
\begin{split}
\sum_{m_b}\mathcal{A}_{ba} &= \int d\Omega\left[\left|\sin(\vartheta)A_{l_b00,10}A_{10,00}\right|^2 + 0 + \left|-\frac{\sqrt{2}}{2}\cos(\vartheta)e^{i\varphi}A_{l_b11,10}A_{10,00}\right|^2 + \left|-i\frac{\sqrt{2}}{2}e^{i\varphi}A_{l_b11,10}A_{10,00}\right|^2 + \right.\\& \hspace{2cm}\left.\left|-\frac{\sqrt{2}}{2}\cos(\vartheta)e^{-i\varphi}A_{l_b-1-1,10}A_{10,00}\right|^2 + \left|i\frac{\sqrt{2}}{2}e^{-i\varphi}A_{l_b-1-1,10}A_{10,00}\right|^2 \right] \\
&= \int d\Omega \left[\sin^2(\vartheta)(A_{l_b00,10}A_{10,00})^2 + \left(\cos^2(\vartheta) + 1\right)(A_{l_b\pm1\pm1, 10}A_{10, 00})^2\right].
\end{split}
\end{align}
\end{widetext}
Integrating over the direction of the scattered photon yields,
\begin{align}
    \sum_{m_b}\mathcal{A}_{ba} = \frac{8\pi}{3}\left((A_{l_b00,10}A_{10,00})^2 + 2(A_{l_b\pm1\pm1, 10}A_{10,00})^2\right).
\end{align}

\subsubsection*{S to S scattering}
When the final state $b$ is also an S state,  $m_b = 0$ only and, 
\begin{align}
    \sum_{m_b}\mathcal{A}_{ba} = \frac{8\pi}{3}\left(\sqrt{\frac{1}{3}}\times\sqrt{\frac{1}{3}}\right)^2 + 0 = \frac{8\pi}{27}.
\end{align}
Rayleigh scattering is a special case of this where $a=b$. Equation (\ref{eqn:rayleigh}) is immediate from these results and equation (\ref{eqn:pol}).

\subsubsection*{S to D scattering}
In the case of S to D scattering, $m_b = \pm 1$ are valid magnetic quantum numbers, hence,
\begin{align}
\begin{split}
    \sum_{m_b}\mathcal{A}_{ba} &= \frac{8\pi}{3}\left[\left(\sqrt{\frac{4}{15}}\times\sqrt{\frac{1}{3}}\right)^2 + 2\left(\sqrt{\frac{2}{10}}\times\sqrt{\frac{1}{3}}\right)^2 \right] \\&= \frac{16\pi}{27}.
    \end{split}
\end{align}

\section{Implicit summation and calculation of radial matrix elements}
\label{app:sturmian}

Calculating the polarisability and scattering rates requires the computation of sums over all atomic states $k$ that are dipole-coupled to both $a$ and $b$. In principle, this includes a sum over an infinite number of discrete, bound states, and an integration across a continuum of unbound states. Following \cite{Potvliege1998, Potvliege1989}, we carry out this calculation using the implicit summation method (also called the Dalgarno-Lewis method \cite{Mavromatis1991, Nandi1996}). Here,  the sum is replaced by a single matrix element:
\begin{align}
\label{eqn:radialelement}
    \sum_k\frac{\boldsymbol{\epsilon}_i\cdot\boldsymbol{r}_{bk}\boldsymbol{\epsilon}_j\cdot\boldsymbol{r}_{ka}}{\omega_{k} - \omega_{a} \mp \omega} = \bra{b}\boldsymbol{\epsilon}_i\cdot\boldsymbol{r}\ket{\Psi},
\end{align}
where the subscripts $i$ and $j$ simply identify the two polarisation vectors, and the vector $\ket{\Psi}$ solves the equation
\begin{align}
\label{eqn:psidefinition}
    [H - (\omega_{a} \pm \omega)]\ket{\Psi} = \boldsymbol{\epsilon}_j\cdot\boldsymbol{r}\ket{a},
\end{align}
where $H$ is the Hamiltonian. The vector $\ket{\Psi}$ includes contributions from both the discrete and continuum parts of the spectrum.

We construct a discrete set of Laguerre functions \cite{hostler1970, Maquet1998}, 
\begin{align}
    \mathcal{B} = \left\{\frac{1}{r}S_{n,l}^{(\zeta)}(r)Y_{lm}(\theta, \phi)\text{ : }n \in \mathbb{N}, l \in \mathbb{N}_0, |m| \leq l\right\},
\end{align}
where $\zeta$ is a free, real parameter. $n$ and $l$ index the Sturmian functions:
\begin{align}
S_{nl}^{(\zeta)}(r) = \mathcal{N}_{nl}(2\zeta r)^{l+1}e^{-\zeta r}L_{n-1}^{2l+1}(2\zeta r),
\end{align}
with $\mathcal{N}_{nl}$ a normalising constant. $L_{x}^{y}(s)$ are the associated Laguerre polynomials, expressed in the Rodriguez representation \cite{Gradshtein2015}.

Such a set of Laguerre functions form a complete set spanning the Hilbert space of $L^2(0, \infty)$ functions --- the space of square integrable functions over the semi-definite interval (see \cite{Szego1939} for proof). Thus, $\mathcal{B}$ forms a complete set spanning the Hilbert space inhabited by radial wave functions of hydrogen. 

We normalise the functions of $\mathcal{B}$ as $\mathcal{N}_{nl} = \sqrt{(n-1)!/(n+2l)!}$ such that, 
\begin{align}
    \int^\infty_0r^2dr \left(\frac{1}{r}S_{n'l}^{(\zeta)}\right)\frac{1}{r}\left(\frac{1}{r}S_{nl}^{(\zeta)}\right) = \delta_{n'n}.
\end{align}
Orthogonality over the indexes $l$ and $m$ is assured by the orthogonality of the spherical harmonics. This condition, and standard relations between the associated Laguerre polynomials as presented in \cite{Gradshtein2015}, can be used to analytically evaluate matrix elements of operators. In this way, we produce  representations of the Hamiltonian dipole operators in the basis $\mathcal{B}$; we denote these matrices $\EuScript{H}$ and $\EuScript{R}$ respectively.

The wave function of an atomic state $a$ is represented as a vector $\boldsymbol{a}$ in the basis $\mathcal{B}$. The angular wave functions of hydrogen are given by the same spherical harmonics that give the angular parts of the basis functions $\mathcal{B}$. Indeed, the quantum numbers $l$ and $m$ relate directly to the indices of the same labels. 
The radial wave function for atomic state $a$ with principal quantum number $N$ can be decomposed over the radial parts of the basis functions:
\begin{align}
    R_{Nl}(r) = \sum_{n\in\mathbb{N}}C_{n, N, l}\frac{1}{r}S_{nl}^{(\zeta)}(r),
\end{align}
where $C_{n, N, l}$ are constants. We calculate this vector by numerically solving the Schr\"{o}dinger equation as a generalised eigenvalue problem:
\begin{align}
    \EuScript{H}\boldsymbol{a} = \omega_a\EuScript{T}\boldsymbol{a}.
\end{align}
The inclusion of an overlap matrix $\EuScript{T}$ accounts for the non-trivial overlap between the Sturmian functions.

The calculation of (\ref{eqn:radialelement}) and (\ref{eqn:psidefinition}) reduces to solving the matrix equation, 
\begin{align}
\label{eqn:matrixeqns1}
    \left[\EuScript{H} - (\omega_a \pm \omega)\EuScript{T}\right]\boldsymbol{\Psi} = \EuScript{R}\boldsymbol{a},
\end{align}
and computing the product,
\begin{align}
\label{eqn:matrixeqns2}
    \boldsymbol{b}^\text{T}\EuScript{R}\boldsymbol{\Psi},
\end{align}
for both sums (i) and (ii) independently.

In general, the set $\mathcal{B}$ is infinite. For computations, it is necessary to restrict the set to a finite basis. We only need to address values of $l$ ranging from 0 to 2, and restrict $n$ to a finite range in $\mathbb{N}_0$. In principle, the choice of $\zeta$ is free, although convergence is much slower for excessively large values of this parameter. For all calculations presented in this text we let $n$ range to 300 and set $\zeta = 0.3$. We find that this is sufficient to ensure that the resulting values of the scattering rate are stable to 1 part in $10^9$ under small variations of $\zeta$.

We use python to perform these calculations. The code is freely available at \cite{Joseph2023code}.

\subsubsection*{A note on calculations.}

\begin{figure}[h!]
    \centering
    \includegraphics[width=0.45\textwidth]{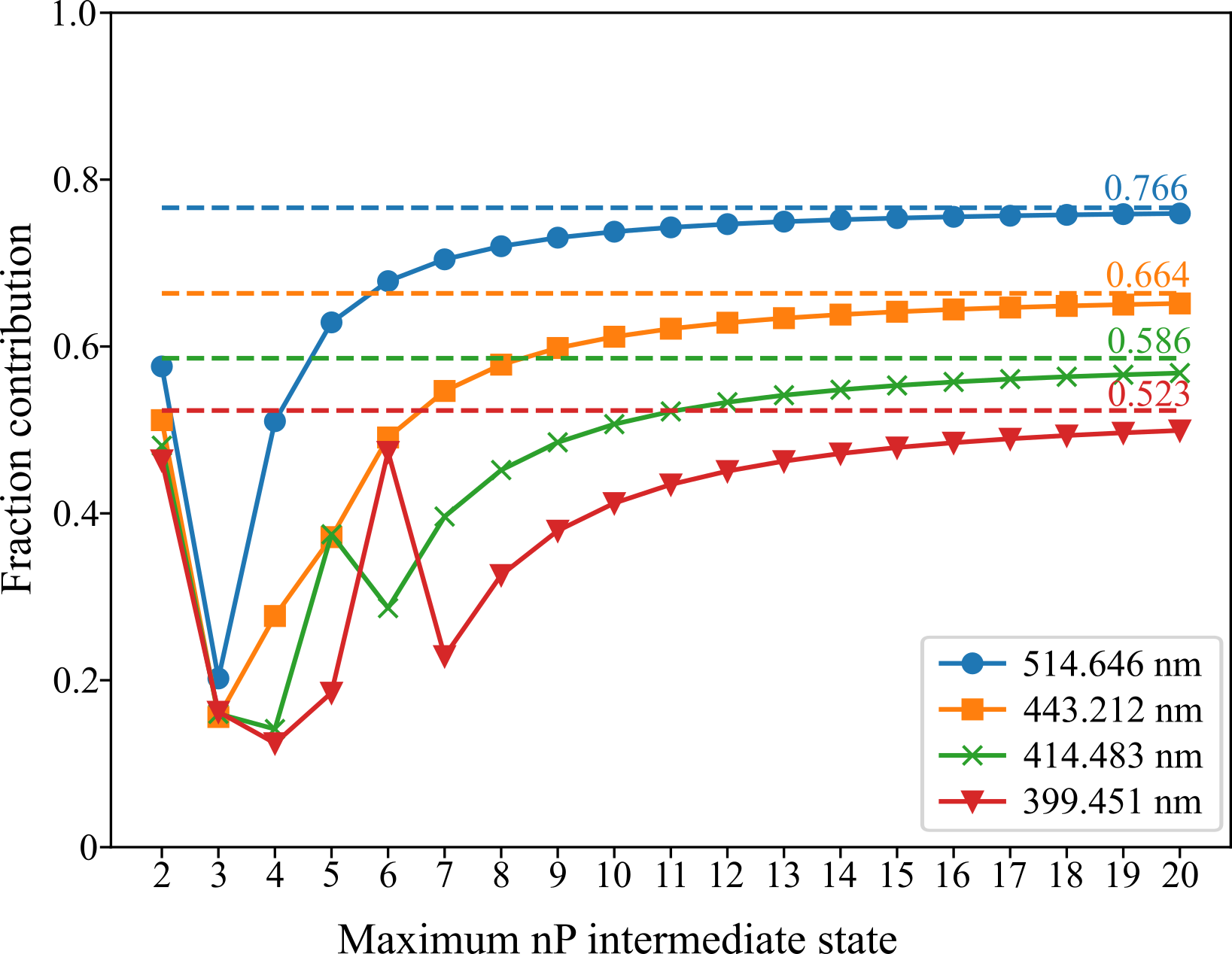}
    \caption{The fraction of the total inelastic scattering rate which can be attributed to the lowest $n$P intermediate state, summed over $n$, for $n$ varying from 2 to 20. The data points and solid lines show how this fraction increases with $n$ at the four magic wavelengths considered in this work. The dashed lines represent the convergent limit of this fraction, calculated for a discrete sum up to the 100P state. The difference between this limit and 1 is the contribution of the continuum states.}
    \label{fig:calculation_contents}
\end{figure}

It is well established that the contributions of continuum states to photon scattering in hydrogen cannot be neglected. There exists a wide literature on analytic (e.g. \cite{Marian1989, Maquet1998,Swainson1991} ) and numerical (e.g. \cite{Potvliege1998, Mcnamara2018, Singor2021}) methods for calculating these matrix elements including both the discrete and continuum parts of the spectrum. For calculations in alkali atoms, it is often sufficient to sum across only a small number of discrete states to converge to the correct value. However, this is not the case in hydrogen. Figure \ref{fig:calculation_contents} illustrates this point. This figure represents the fraction of the total inelastic scattering rate that is obtained by summing over a finite number of bound intermediate states only. As seen from this figure, failing to consider the continuum when calculating these scattering rates would underestimate them by 23--48\% of their true value at the magic wavelengths, with the underestimation increasing with photon energy.

\section{Doppler free spectroscopy in optical traps}
\label{app:LDregime}

Atoms recoil under absorption of a photon, as linear momentum is transferred from the photon to the atom. For atoms confined in optical traps, this manifests as a change in vibrational trap state. The matrix element for transitions from vibrational state $n$ to $n'$ is proportional to $\bra{n'}e^{i\boldsymbol{k}\cdot\boldsymbol{r}}\ket{n}$ \cite{eschner2003}.
In the case of Doppler free excitation on a two photon transition, the atom experiences two separate momentum kicks in opposing directions.
Assuming that the photons come from opposing beams oriented along the lattice axis ($\hat{\boldsymbol{x}})$, the atom experiences two momentum kicks: $+kx$ and $-kx$. This results in the matrix elements:
\begin{align}
    \bra{n'}e^{ikx}+e^{-ikx}\ket{n} = \bra{n'}e^{i\eta(a^\dag+a)}+e^{-i\eta(a^\dag+a)}\ket{n},
    \label{eqn:vibelms}
\end{align}
where we have approximated the potential at bottom of a given lattice site as harmonic: $a$ and $a^\dag$ are the ladder operators for the simple harmonic oscillator. The Lamb Dicke parameter $\eta = \sqrt{E_\text{probe}/\omega_\text{T}}$ is the ratio between the recoil energy associated with probe photon absorption and the separation of vibrational states.

We can expand the matrix element \ref{eqn:vibelms} in powers of $i\eta$. When $\eta$ is small, higher order terms can be neglected and the matrix element can be written:
\begin{align}
    \bra{n'} 2 + 0 - \eta^2(a^\dag+a)^2 + 0 +O(\eta^4)\ket{n}.
\end{align}
Terms with odd powers $i\eta$ and $-i\eta$ cancel, while terms with even powers remain. The cancellation of the first order terms is analogous to the elimination of the Doppler shift in field free measurements and here relates to the suppression of first order sidebands.

Constraining consideration to leading order, one finds the probability of transition between vibrational states:
\begin{align}
    P_{n\rightarrow n} = \frac{1}{N}(2 - \eta^2 - 2n\eta^2)^2\\
    P_{n\rightarrow n+2} = \frac{1}{N}\eta^4(n+1)(n+2)\\
    P_{n\rightarrow n-2} = \frac{1}{N}\eta^4(n)(n-1),
\end{align}
where $N$ is some normalising function such that $\sum_{n'} P_{n \rightarrow n'} = 1$ for all vibrational states $n'$. Transitions $n \rightarrow n$ contribute to the central carrier signal, whilst transitions $n \rightarrow n\pm2$ contribute to second order sidebands and are detuned from the carrier by $\pm 2\nu_\text{T}$.

\subsection*{The second order Doppler shift}
Following the arguments of \cite{martinez2022}, one obtains the following relativistic shift to the probe frequency in the atom frame:
\begin{align}
    \left\langle\frac{\delta \hat{\nu}}{\nu}\right\rangle_n = -\frac{h\nu_\text{T}}{4m_\text{H}c^2}(2n + 1) - \left(\frac{g}{2\pi\nu_\text{T}c}\right)^2 + \frac{\phi_0}{c^2}.
\end{align}
where $g$ is the acceleration due to gravity in the measurement direction and $\phi_0$ is the gravitational potential at the centre of a given lattice site.
The first term is the frequency shift from the second-order Doppler (SOD) effect whilst the second and third terms describe gravitational red-shift. The gravitational red-shift terms are negligible compared to the SOD term, $O(10^{-29})$ compared to $O(10^{-17}$), so we have:
\begin{align}
    \left\langle\frac{\delta \hat{\nu}}{\nu}\right\rangle_n = -\frac{h\nu_\text{T}}{4m_\text{H}c^2}(2n + 1).
    \label{eqn:SODonly}
\end{align}
Consider frequency shifts in an optical potential with trap frequency $8.79$\,MHz (as in table \ref{tab:LDdepths}). The SOD shift terms depend upon vibrational state. For $n=0$ we obtain a fractional shift of $-9.68\times10^{-18}$. 

The shift in a thermal state at temperature $T$ can be found by replacing $n$ in equation (\ref{eqn:SODonly}) with the average occupation number:
\begin{align}
    \bar{n} = \frac{1}{\exp\left(\frac{h\nu_\text{T}}{k_\text{B}T}\right) - 1},
\end{align}
where $k_\text{B}$ is the Boltzmann constant. 

The lattices described in table \ref{tab:LDdepths} have depths $|U_0|/k_\mathrm{B}$ $O(1)$\,mK. Table \ref{tab:thermal_shifts} contains the SOD shift for thermal states with temperature $T=|U_0|/k_\mathrm{B}$ and $T=|U_0|/(3 k_\mathrm{B})$ in each magic wavelength lattice. The shallow traps and large trap frequencies lead to low average occupations $\bar{n}$, so thermal state shifts remain small.

\begin{table}[h!]
\centering
\caption{Fractional SOD shift $\braket{\delta\hat{\nu}/\nu}$ for thermal states in magic wavelength lattices with trap frequencies of $8.79$\,MHz.}
\label{tab:thermal_shifts}
    \begin{tabular}{c | c c c c c}
     \hline\hline \\ [-0.15cm]
        $\lambda$\,/nm & $\frac{|U_0|}{k_\mathrm{B}}$\,/mK & $\bar{n}_{T=|U_0|/{k_\mathrm{B}}}$ & $\braket{\delta\hat{\nu}/\nu}_{T=|U_0|/{k_\mathrm{B}}}$ &  $\bar{n}_{T=|U_0|/(3 {k_\mathrm{B}})}$ & $\braket{\delta\hat{\nu}/\nu}_{T=|U_0|/(3 {k_\mathrm{B}})}$\\[0.2cm]
        \hline \\ [-0.15cm]
        514.6 & 1.24 & 2.47 & $-5.75\times10^{-17}$ & 0.563 & $-2.06\times10^{-17}$\\[0.1cm]
        443.2 & 0.919 & 1.72 & $-4.29\times10^{-17}$ & 0.337 & $-1.62\times10^{-17}$\\[0.1cm]
        414.5 & 0.808 & 1.46 & $-3.79\times10^{-17}$ & 0.264 & $-1.48\times10^{-17}$\\[0.1cm]
        399.5 & 0.744 & 1.31 & $-3.51\times10^{-17}$ & 0.223 & $-1.40\times10^{-17}$\\ [0.2cm]
        \hline
    \end{tabular}
\end{table}

\bibliography{Bibliography}

\providecommand{\noopsort}[1]{}\providecommand{\singleletter}[1]{#1}%
\begin{thebibliography}{85}%
\makeatletter
\providecommand \@ifxundefined [1]{%
 \@ifx{#1\undefined}
}%
\providecommand \@ifnum [1]{%
 \ifnum #1\expandafter \@firstoftwo
 \else \expandafter \@secondoftwo
 \fi
}%
\providecommand \@ifx [1]{%
 \ifx #1\expandafter \@firstoftwo
 \else \expandafter \@secondoftwo
 \fi
}%
\providecommand \natexlab [1]{#1}%
\providecommand \enquote  [1]{``#1''}%
\providecommand \bibnamefont  [1]{#1}%
\providecommand \bibfnamefont [1]{#1}%
\providecommand \citenamefont [1]{#1}%
\providecommand \href@noop [0]{\@secondoftwo}%
\providecommand \href [0]{\begingroup \@sanitize@url \@href}%
\providecommand \@href[1]{\@@startlink{#1}\@@href}%
\providecommand \@@href[1]{\endgroup#1\@@endlink}%
\providecommand \@sanitize@url [0]{\catcode `\\12\catcode `\$12\catcode
  `\&12\catcode `\#12\catcode `\^12\catcode `\_12\catcode `\%12\relax}%
\providecommand \@@startlink[1]{}%
\providecommand \@@endlink[0]{}%
\providecommand \url  [0]{\begingroup\@sanitize@url \@url }%
\providecommand \@url [1]{\endgroup\@href {#1}{\urlprefix }}%
\providecommand \urlprefix  [0]{URL }%
\providecommand \Eprint [0]{\href }%
\providecommand \doibase [0]{https://doi.org/}%
\providecommand \selectlanguage [0]{\@gobble}%
\providecommand \bibinfo  [0]{\@secondoftwo}%
\providecommand \bibfield  [0]{\@secondoftwo}%
\providecommand \translation [1]{[#1]}%
\providecommand \BibitemOpen [0]{}%
\providecommand \bibitemStop [0]{}%
\providecommand \bibitemNoStop [0]{.\EOS\space}%
\providecommand \EOS [0]{\spacefactor3000\relax}%
\providecommand \BibitemShut  [1]{\csname bibitem#1\endcsname}%
\let\auto@bib@innerbib\@empty
\bibitem [{\citenamefont {Horbatsch}\ and\ \citenamefont
  {Hessels}(2016)}]{Horbatsch2016}%
  \BibitemOpen
  \bibfield  {author} {\bibinfo {author} {\bibfnamefont {M.}~\bibnamefont
  {Horbatsch}}\ and\ \bibinfo {author} {\bibfnamefont {E.~A.}\ \bibnamefont
  {Hessels}},\ }\href {https://doi.org/10.1103/PhysRevA.93.022513} {\bibfield
  {journal} {\bibinfo  {journal} {Physical Review A}\ }\textbf {\bibinfo
  {volume} {93}},\ \bibinfo {pages} {022513} (\bibinfo {year}
  {2016})}\BibitemShut {NoStop}%
\bibitem [{\citenamefont {H{\"a}nsch}(2006)}]{hansch2006}%
  \BibitemOpen
  \bibfield  {author} {\bibinfo {author} {\bibfnamefont {T.~W.}\ \bibnamefont
  {H{\"a}nsch}},\ }\href@noop {} {\bibfield  {journal} {\bibinfo  {journal}
  {Reviews of Modern Physics}\ }\textbf {\bibinfo {volume} {78}},\ \bibinfo
  {pages} {1297} (\bibinfo {year} {2006})}\BibitemShut {NoStop}%
\bibitem [{\citenamefont {Jones}\ \emph {et~al.}(2020)\citenamefont {Jones},
  \citenamefont {Potvliege},\ and\ \citenamefont {Spannowsky}}]{Jones2020}%
  \BibitemOpen
  \bibfield  {author} {\bibinfo {author} {\bibfnamefont {M.~P.}\ \bibnamefont
  {Jones}}, \bibinfo {author} {\bibfnamefont {R.~M.}\ \bibnamefont
  {Potvliege}},\ and\ \bibinfo {author} {\bibfnamefont {M.}~\bibnamefont
  {Spannowsky}},\ }\href {https://doi.org/10.1103/PhysRevResearch.2.013244}
  {\bibfield  {journal} {\bibinfo  {journal} {Physical Review Research}\
  }\textbf {\bibinfo {volume} {2}},\ \bibinfo {pages} {013244} (\bibinfo {year}
  {2020})}\BibitemShut {NoStop}%
\bibitem [{\citenamefont {Safronova}\ \emph {et~al.}(2018)\citenamefont
  {Safronova}, \citenamefont {Budker}, \citenamefont {DeMille}, \citenamefont
  {Kimball}, \citenamefont {Derevianko},\ and\ \citenamefont
  {Clark}}]{Safronova2018}%
  \BibitemOpen
  \bibfield  {author} {\bibinfo {author} {\bibfnamefont {M.}~\bibnamefont
  {Safronova}}, \bibinfo {author} {\bibfnamefont {D.}~\bibnamefont {Budker}},
  \bibinfo {author} {\bibfnamefont {D.}~\bibnamefont {DeMille}}, \bibinfo
  {author} {\bibfnamefont {D.~F.~J.}\ \bibnamefont {Kimball}}, \bibinfo
  {author} {\bibfnamefont {A.}~\bibnamefont {Derevianko}},\ and\ \bibinfo
  {author} {\bibfnamefont {C.~W.}\ \bibnamefont {Clark}},\ }\href
  {https://doi.org/10.1103/RevModPhys.90.025008} {\bibfield  {journal}
  {\bibinfo  {journal} {Reviews of Modern Physics}\ }\textbf {\bibinfo {volume}
  {90}},\ \bibinfo {pages} {025008} (\bibinfo {year} {2018})}\BibitemShut
  {NoStop}%
\bibitem [{\citenamefont {Karshenboim}(2010)}]{Karshenboim2010}%
  \BibitemOpen
  \bibfield  {author} {\bibinfo {author} {\bibfnamefont {S.~G.}\ \bibnamefont
  {Karshenboim}},\ }\href {https://doi.org/10.1103/PhysRevLett.104.220406}
  {\bibfield  {journal} {\bibinfo  {journal} {Physical review letters}\
  }\textbf {\bibinfo {volume} {104}},\ \bibinfo {pages} {220406} (\bibinfo
  {year} {2010})}\BibitemShut {NoStop}%
\bibitem [{\citenamefont {Karshenboim}(2005)}]{Karshenboim2005}%
  \BibitemOpen
  \bibfield  {author} {\bibinfo {author} {\bibfnamefont {S.~G.}\ \bibnamefont
  {Karshenboim}},\ }\href {https://doi.org/10.1016/j.physrep.2005.08.008}
  {\bibfield  {journal} {\bibinfo  {journal} {Physics reports}\ }\textbf
  {\bibinfo {volume} {422}},\ \bibinfo {pages} {1} (\bibinfo {year}
  {2005})}\BibitemShut {NoStop}%
\bibitem [{\citenamefont {Brax}\ and\ \citenamefont
  {Burrage}(2011)}]{Brax2011}%
  \BibitemOpen
  \bibfield  {author} {\bibinfo {author} {\bibfnamefont {P.}~\bibnamefont
  {Brax}}\ and\ \bibinfo {author} {\bibfnamefont {C.}~\bibnamefont {Burrage}},\
  }\href {https://doi.org/10.1103/PhysRevD.83.035020} {\bibfield  {journal}
  {\bibinfo  {journal} {Physical Review D}\ }\textbf {\bibinfo {volume} {83}},\
  \bibinfo {pages} {035020} (\bibinfo {year} {2011})}\BibitemShut {NoStop}%
\bibitem [{\citenamefont {Burrage}\ and\ \citenamefont
  {Sakstein}(2018)}]{Burrage2018}%
  \BibitemOpen
  \bibfield  {author} {\bibinfo {author} {\bibfnamefont {C.}~\bibnamefont
  {Burrage}}\ and\ \bibinfo {author} {\bibfnamefont {J.}~\bibnamefont
  {Sakstein}},\ }\href {https://doi.org/10.1007/s41114-018-0011} {\bibfield
  {journal} {\bibinfo  {journal} {Living reviews in relativity}\ }\textbf
  {\bibinfo {volume} {21}},\ \bibinfo {pages} {1} (\bibinfo {year}
  {2018})}\BibitemShut {NoStop}%
\bibitem [{\citenamefont {Stadnik}(2018)}]{Stadnik2018}%
  \BibitemOpen
  \bibfield  {author} {\bibinfo {author} {\bibfnamefont {Y.~V.}\ \bibnamefont
  {Stadnik}},\ }\href {https://doi.org/10.1103/PhysRevLett.120.223202}
  {\bibfield  {journal} {\bibinfo  {journal} {Physical Review Letters}\
  }\textbf {\bibinfo {volume} {120}},\ \bibinfo {pages} {223202} (\bibinfo
  {year} {2018})}\BibitemShut {NoStop}%
\bibitem [{\citenamefont {Ghosh}\ \emph {et~al.}(2020)\citenamefont {Ghosh},
  \citenamefont {Grossman},\ and\ \citenamefont {Tangarife}}]{Ghosh2020}%
  \BibitemOpen
  \bibfield  {author} {\bibinfo {author} {\bibfnamefont {M.}~\bibnamefont
  {Ghosh}}, \bibinfo {author} {\bibfnamefont {Y.}~\bibnamefont {Grossman}},\
  and\ \bibinfo {author} {\bibfnamefont {W.}~\bibnamefont {Tangarife}},\ }\href
  {https://doi.org/10.1103/PhysRevD.101.116006} {\bibfield  {journal} {\bibinfo
   {journal} {Physical Review D}\ }\textbf {\bibinfo {volume} {101}},\ \bibinfo
  {pages} {116006} (\bibinfo {year} {2020})}\BibitemShut {NoStop}%
\bibitem [{\citenamefont {Delaunay}\ \emph {et~al.}(2023)\citenamefont
  {Delaunay}, \citenamefont {Karr}, \citenamefont {Kitahara}, \citenamefont
  {Koelemeij}, \citenamefont {Soreq},\ and\ \citenamefont
  {Zupan}}]{delaunay2023}%
  \BibitemOpen
  \bibfield  {author} {\bibinfo {author} {\bibfnamefont {C.}~\bibnamefont
  {Delaunay}}, \bibinfo {author} {\bibfnamefont {J.-P.}\ \bibnamefont {Karr}},
  \bibinfo {author} {\bibfnamefont {T.}~\bibnamefont {Kitahara}}, \bibinfo
  {author} {\bibfnamefont {J.~C.}\ \bibnamefont {Koelemeij}}, \bibinfo {author}
  {\bibfnamefont {Y.}~\bibnamefont {Soreq}},\ and\ \bibinfo {author}
  {\bibfnamefont {J.}~\bibnamefont {Zupan}},\ }\href@noop {} {\bibfield
  {journal} {\bibinfo  {journal} {Physical review letters}\ }\textbf {\bibinfo
  {volume} {130}},\ \bibinfo {pages} {121801} (\bibinfo {year}
  {2023})}\BibitemShut {NoStop}%
\bibitem [{\citenamefont {Shore}(2005)}]{Shore2005}%
  \BibitemOpen
  \bibfield  {author} {\bibinfo {author} {\bibfnamefont {G.~M.}\ \bibnamefont
  {Shore}},\ }\href {https://doi.org/10.1016/j.nuclphysb.2005.03.040.}
  {\bibfield  {journal} {\bibinfo  {journal} {J. Nucl. Phys. B}\ }\textbf
  {\bibinfo {volume} {717}},\ \bibinfo {pages} {86} (\bibinfo {year}
  {2005})}\BibitemShut {NoStop}%
\bibitem [{\citenamefont {Kosteleck{\`y}}\ and\ \citenamefont
  {Vargas}(2015)}]{Kostelecky2015}%
  \BibitemOpen
  \bibfield  {author} {\bibinfo {author} {\bibfnamefont {V.~A.}\ \bibnamefont
  {Kosteleck{\`y}}}\ and\ \bibinfo {author} {\bibfnamefont {A.~J.}\
  \bibnamefont {Vargas}},\ }\href {https://doi.org/10.1103/PhysRevD.92.056002}
  {\bibfield  {journal} {\bibinfo  {journal} {Physical Review D}\ }\textbf
  {\bibinfo {volume} {92}},\ \bibinfo {pages} {056002} (\bibinfo {year}
  {2015})}\BibitemShut {NoStop}%
\bibitem [{\citenamefont {Ahmadi}\ \emph {et~al.}(2017)\citenamefont {Ahmadi},
  \citenamefont {Alves}, \citenamefont {Baker}, \citenamefont {Bertsche},
  \citenamefont {Butler}, \citenamefont {Capra}, \citenamefont {Carruth},
  \citenamefont {Cesar}, \citenamefont {Charlton}, \citenamefont {Cohen} \emph
  {et~al.}}]{Ahmadi2017}%
  \BibitemOpen
  \bibfield  {author} {\bibinfo {author} {\bibfnamefont {M.}~\bibnamefont
  {Ahmadi}}, \bibinfo {author} {\bibfnamefont {B.~X.~R.}\ \bibnamefont
  {Alves}}, \bibinfo {author} {\bibfnamefont {C.}~\bibnamefont {Baker}},
  \bibinfo {author} {\bibfnamefont {W.}~\bibnamefont {Bertsche}}, \bibinfo
  {author} {\bibfnamefont {E.}~\bibnamefont {Butler}}, \bibinfo {author}
  {\bibfnamefont {A.}~\bibnamefont {Capra}}, \bibinfo {author} {\bibfnamefont
  {C.}~\bibnamefont {Carruth}}, \bibinfo {author} {\bibfnamefont
  {C.}~\bibnamefont {Cesar}}, \bibinfo {author} {\bibfnamefont
  {M.}~\bibnamefont {Charlton}}, \bibinfo {author} {\bibfnamefont
  {S.}~\bibnamefont {Cohen}}, \emph {et~al.},\ }\href
  {https://doi.org/10.1038/nature21040} {\bibfield  {journal} {\bibinfo
  {journal} {Nature}\ }\textbf {\bibinfo {volume} {541}},\ \bibinfo {pages}
  {506} (\bibinfo {year} {2017})}\BibitemShut {NoStop}%
\bibitem [{\citenamefont {Ahmadi}\ \emph {et~al.}(2018)\citenamefont {Ahmadi},
  \citenamefont {Alves}, \citenamefont {Baker}, \citenamefont {Bertsche},
  \citenamefont {Capra}, \citenamefont {Carruth}, \citenamefont {Cesar},
  \citenamefont {Charlton}, \citenamefont {Cohen}, \citenamefont {Collister}
  \emph {et~al.}}]{Ahmadi2018}%
  \BibitemOpen
  \bibfield  {author} {\bibinfo {author} {\bibfnamefont {M.}~\bibnamefont
  {Ahmadi}}, \bibinfo {author} {\bibfnamefont {B.}~\bibnamefont {Alves}},
  \bibinfo {author} {\bibfnamefont {C.}~\bibnamefont {Baker}}, \bibinfo
  {author} {\bibfnamefont {W.}~\bibnamefont {Bertsche}}, \bibinfo {author}
  {\bibfnamefont {A.}~\bibnamefont {Capra}}, \bibinfo {author} {\bibfnamefont
  {C.}~\bibnamefont {Carruth}}, \bibinfo {author} {\bibfnamefont
  {C.}~\bibnamefont {Cesar}}, \bibinfo {author} {\bibfnamefont
  {M.}~\bibnamefont {Charlton}}, \bibinfo {author} {\bibfnamefont
  {S.}~\bibnamefont {Cohen}}, \bibinfo {author} {\bibfnamefont
  {R.}~\bibnamefont {Collister}}, \emph {et~al.},\ }\href
  {https://doi.org/10.1038/s41586-018-0017-2} {\bibfield  {journal} {\bibinfo
  {journal} {Nature}\ }\textbf {\bibinfo {volume} {557}},\ \bibinfo {pages}
  {71} (\bibinfo {year} {2018})}\BibitemShut {NoStop}%
\bibitem [{\citenamefont {Alexandrou}\ \emph {et~al.}(2020)\citenamefont
  {Alexandrou}, \citenamefont {Hadjiyiannakou}, \citenamefont {Koutsou},
  \citenamefont {Ottnad},\ and\ \citenamefont {Petschlies}}]{Alexandrou2020}%
  \BibitemOpen
  \bibfield  {author} {\bibinfo {author} {\bibfnamefont {C.}~\bibnamefont
  {Alexandrou}}, \bibinfo {author} {\bibfnamefont {K.}~\bibnamefont
  {Hadjiyiannakou}}, \bibinfo {author} {\bibfnamefont {G.}~\bibnamefont
  {Koutsou}}, \bibinfo {author} {\bibfnamefont {K.}~\bibnamefont {Ottnad}},\
  and\ \bibinfo {author} {\bibfnamefont {M.}~\bibnamefont {Petschlies}},\
  }\href {https://doi.org/10.1103/PhysRevD.101.114504} {\bibfield  {journal}
  {\bibinfo  {journal} {Physical Review D}\ }\textbf {\bibinfo {volume}
  {101}},\ \bibinfo {pages} {114504} (\bibinfo {year} {2020})}\BibitemShut
  {NoStop}%
\bibitem [{\citenamefont {Parthey}\ \emph {et~al.}(2011)\citenamefont
  {Parthey}, \citenamefont {Matveev}, \citenamefont {Alnis}, \citenamefont
  {Bernhardt}, \citenamefont {Beyer}, \citenamefont {Holzwarth}, \citenamefont
  {Maistrou}, \citenamefont {Pohl}, \citenamefont {Predehl}, \citenamefont
  {Udem}, \citenamefont {Wilken}, \citenamefont {Kolachevsky}, \citenamefont
  {Abgrall}, \citenamefont {Rovera}, \citenamefont {Salomon}, \citenamefont
  {Laurent},\ and\ \citenamefont {H\"ansch}}]{Parthey2011}%
  \BibitemOpen
  \bibfield  {author} {\bibinfo {author} {\bibfnamefont {C.~G.}\ \bibnamefont
  {Parthey}}, \bibinfo {author} {\bibfnamefont {A.}~\bibnamefont {Matveev}},
  \bibinfo {author} {\bibfnamefont {J.}~\bibnamefont {Alnis}}, \bibinfo
  {author} {\bibfnamefont {B.}~\bibnamefont {Bernhardt}}, \bibinfo {author}
  {\bibfnamefont {A.}~\bibnamefont {Beyer}}, \bibinfo {author} {\bibfnamefont
  {R.}~\bibnamefont {Holzwarth}}, \bibinfo {author} {\bibfnamefont
  {A.}~\bibnamefont {Maistrou}}, \bibinfo {author} {\bibfnamefont
  {R.}~\bibnamefont {Pohl}}, \bibinfo {author} {\bibfnamefont {K.}~\bibnamefont
  {Predehl}}, \bibinfo {author} {\bibfnamefont {T.}~\bibnamefont {Udem}},
  \bibinfo {author} {\bibfnamefont {T.}~\bibnamefont {Wilken}}, \bibinfo
  {author} {\bibfnamefont {N.}~\bibnamefont {Kolachevsky}}, \bibinfo {author}
  {\bibfnamefont {M.}~\bibnamefont {Abgrall}}, \bibinfo {author} {\bibfnamefont
  {D.}~\bibnamefont {Rovera}}, \bibinfo {author} {\bibfnamefont
  {C.}~\bibnamefont {Salomon}}, \bibinfo {author} {\bibfnamefont
  {P.}~\bibnamefont {Laurent}},\ and\ \bibinfo {author} {\bibfnamefont {T.~W.}\
  \bibnamefont {H\"ansch}},\ }\href
  {https://doi.org/10.1103/PhysRevLett.107.203001} {\bibfield  {journal}
  {\bibinfo  {journal} {Physical Review Letters}\ }\textbf {\bibinfo {volume}
  {107}},\ \bibinfo {pages} {203001} (\bibinfo {year} {2011})}\BibitemShut
  {NoStop}%
\bibitem [{\citenamefont {Matveev}\ \emph {et~al.}(2013)\citenamefont
  {Matveev}, \citenamefont {Parthey}, \citenamefont {Predehl}, \citenamefont
  {Alnis}, \citenamefont {Beyer}, \citenamefont {Holzwarth}, \citenamefont
  {Udem}, \citenamefont {Wilken}, \citenamefont {Kolachevsky}, \citenamefont
  {Abgrall}, \citenamefont {Rovera}, \citenamefont {Salomon}, \citenamefont
  {Laurent}, \citenamefont {Grosche}, \citenamefont {Terra}, \citenamefont
  {Legero}, \citenamefont {Schnatz}, \citenamefont {Weyers}, \citenamefont
  {Altschul},\ and\ \citenamefont {H\"ansch}}]{Matveev2013}%
  \BibitemOpen
  \bibfield  {author} {\bibinfo {author} {\bibfnamefont {A.}~\bibnamefont
  {Matveev}}, \bibinfo {author} {\bibfnamefont {C.~G.}\ \bibnamefont
  {Parthey}}, \bibinfo {author} {\bibfnamefont {K.}~\bibnamefont {Predehl}},
  \bibinfo {author} {\bibfnamefont {J.}~\bibnamefont {Alnis}}, \bibinfo
  {author} {\bibfnamefont {A.}~\bibnamefont {Beyer}}, \bibinfo {author}
  {\bibfnamefont {R.}~\bibnamefont {Holzwarth}}, \bibinfo {author}
  {\bibfnamefont {T.}~\bibnamefont {Udem}}, \bibinfo {author} {\bibfnamefont
  {T.}~\bibnamefont {Wilken}}, \bibinfo {author} {\bibfnamefont
  {N.}~\bibnamefont {Kolachevsky}}, \bibinfo {author} {\bibfnamefont
  {M.}~\bibnamefont {Abgrall}}, \bibinfo {author} {\bibfnamefont
  {D.}~\bibnamefont {Rovera}}, \bibinfo {author} {\bibfnamefont
  {C.}~\bibnamefont {Salomon}}, \bibinfo {author} {\bibfnamefont
  {P.}~\bibnamefont {Laurent}}, \bibinfo {author} {\bibfnamefont
  {G.}~\bibnamefont {Grosche}}, \bibinfo {author} {\bibfnamefont
  {O.}~\bibnamefont {Terra}}, \bibinfo {author} {\bibfnamefont
  {T.}~\bibnamefont {Legero}}, \bibinfo {author} {\bibfnamefont
  {H.}~\bibnamefont {Schnatz}}, \bibinfo {author} {\bibfnamefont
  {S.}~\bibnamefont {Weyers}}, \bibinfo {author} {\bibfnamefont
  {B.}~\bibnamefont {Altschul}},\ and\ \bibinfo {author} {\bibfnamefont
  {T.~W.}\ \bibnamefont {H\"ansch}},\ }\href
  {https://doi.org/10.1103/PhysRevLett.110.230801} {\bibfield  {journal}
  {\bibinfo  {journal} {Physical Review Letters}\ }\textbf {\bibinfo {volume}
  {110}},\ \bibinfo {pages} {230801} (\bibinfo {year} {2013})}\BibitemShut
  {NoStop}%
\bibitem [{\citenamefont {Xiong}\ and\ \citenamefont {Peng}(2023)}]{Xiong2023}%
  \BibitemOpen
  \bibfield  {author} {\bibinfo {author} {\bibfnamefont {W.}~\bibnamefont
  {Xiong}}\ and\ \bibinfo {author} {\bibfnamefont {C.}~\bibnamefont {Peng}},\
  }\href {https://doi.org/10.3390/universe9040182} {\bibfield  {journal}
  {\bibinfo  {journal} {Universe}\ }\textbf {\bibinfo {volume} {9}},\ \bibinfo
  {pages} {182} (\bibinfo {year} {2023})}\BibitemShut {NoStop}%
\bibitem [{\citenamefont {Tiesinga}\ \emph {et~al.}(2021)\citenamefont
  {Tiesinga}, \citenamefont {Mohr}, \citenamefont {Newell},\ and\ \citenamefont
  {Taylor}}]{Tiesinga2018}%
  \BibitemOpen
  \bibfield  {author} {\bibinfo {author} {\bibfnamefont {E.}~\bibnamefont
  {Tiesinga}}, \bibinfo {author} {\bibfnamefont {P.~J.}\ \bibnamefont {Mohr}},
  \bibinfo {author} {\bibfnamefont {D.~B.}\ \bibnamefont {Newell}},\ and\
  \bibinfo {author} {\bibfnamefont {B.~N.}\ \bibnamefont {Taylor}},\ }\href
  {https://doi.org/10.1103/RevModPhys.93.025010} {\bibfield  {journal}
  {\bibinfo  {journal} {Rev. Mod. Phys.}\ }\textbf {\bibinfo {volume} {93}},\
  \bibinfo {pages} {025010} (\bibinfo {year} {2021})}\BibitemShut {NoStop}%
\bibitem [{\citenamefont {Mohr}\ \emph {et~al.}(2016)\citenamefont {Mohr},
  \citenamefont {Newell},\ and\ \citenamefont {Taylor}}]{Mohr2014}%
  \BibitemOpen
  \bibfield  {author} {\bibinfo {author} {\bibfnamefont {P.~J.}\ \bibnamefont
  {Mohr}}, \bibinfo {author} {\bibfnamefont {D.~B.}\ \bibnamefont {Newell}},\
  and\ \bibinfo {author} {\bibfnamefont {B.~N.}\ \bibnamefont {Taylor}},\
  }\href {https://doi.org/10.1103/RevModPhys.88.035009} {\bibfield  {journal}
  {\bibinfo  {journal} {Journal of Physical and Chemical Reference Data}\
  }\textbf {\bibinfo {volume} {45}},\ \bibinfo {pages} {043102} (\bibinfo
  {year} {2016})}\BibitemShut {NoStop}%
\bibitem [{\citenamefont {Pohl}\ \emph {et~al.}(2013)\citenamefont {Pohl},
  \citenamefont {Gilman}, \citenamefont {Miller},\ and\ \citenamefont
  {Pachucki}}]{Pohl2013}%
  \BibitemOpen
  \bibfield  {author} {\bibinfo {author} {\bibfnamefont {R.}~\bibnamefont
  {Pohl}}, \bibinfo {author} {\bibfnamefont {R.}~\bibnamefont {Gilman}},
  \bibinfo {author} {\bibfnamefont {G.~A.}\ \bibnamefont {Miller}},\ and\
  \bibinfo {author} {\bibfnamefont {K.}~\bibnamefont {Pachucki}},\ }\href
  {https://doi.org/10.1146/annurev-nucl-102212-170627} {\bibfield  {journal}
  {\bibinfo  {journal} {Annual Review of Nuclear and Particle Science}\
  }\textbf {\bibinfo {volume} {63}},\ \bibinfo {pages} {175} (\bibinfo {year}
  {2013})}\BibitemShut {NoStop}%
\bibitem [{\citenamefont {Gao}\ and\ \citenamefont
  {Vanderhaeghen}(2022)}]{Gao2022}%
  \BibitemOpen
  \bibfield  {author} {\bibinfo {author} {\bibfnamefont {H.}~\bibnamefont
  {Gao}}\ and\ \bibinfo {author} {\bibfnamefont {M.}~\bibnamefont
  {Vanderhaeghen}},\ }\href {https://doi.org/10.1103/RevModPhys.94.015002}
  {\bibfield  {journal} {\bibinfo  {journal} {Reviews of Modern Physics}\
  }\textbf {\bibinfo {volume} {94}},\ \bibinfo {pages} {015002} (\bibinfo
  {year} {2022})}\BibitemShut {NoStop}%
\bibitem [{\citenamefont {Brandt}\ \emph {et~al.}(2022)\citenamefont {Brandt},
  \citenamefont {Cooper}, \citenamefont {Rasor}, \citenamefont {Burkley},
  \citenamefont {Matveev},\ and\ \citenamefont {Yost}}]{Yost2022}%
  \BibitemOpen
  \bibfield  {author} {\bibinfo {author} {\bibfnamefont {A.~D.}\ \bibnamefont
  {Brandt}}, \bibinfo {author} {\bibfnamefont {S.~F.}\ \bibnamefont {Cooper}},
  \bibinfo {author} {\bibfnamefont {C.}~\bibnamefont {Rasor}}, \bibinfo
  {author} {\bibfnamefont {Z.}~\bibnamefont {Burkley}}, \bibinfo {author}
  {\bibfnamefont {A.}~\bibnamefont {Matveev}},\ and\ \bibinfo {author}
  {\bibfnamefont {D.~C.}\ \bibnamefont {Yost}},\ }\href
  {https://doi.org/10.1103/PhysRevLett.128.023001} {\bibfield  {journal}
  {\bibinfo  {journal} {Physical Review Letters}\ }\textbf {\bibinfo {volume}
  {128}},\ \bibinfo {pages} {023001} (\bibinfo {year} {2022})}\BibitemShut
  {NoStop}%
\bibitem [{\citenamefont {Fleurbaey}\ \emph {et~al.}(2018)\citenamefont
  {Fleurbaey}, \citenamefont {Galtier}, \citenamefont {Thomas}, \citenamefont
  {Bonnaud}, \citenamefont {Julien}, \citenamefont {Biraben}, \citenamefont
  {Nez}, \citenamefont {Abgrall},\ and\ \citenamefont
  {Gu\'ena}}]{Fleurbaey2018}%
  \BibitemOpen
  \bibfield  {author} {\bibinfo {author} {\bibfnamefont {H.}~\bibnamefont
  {Fleurbaey}}, \bibinfo {author} {\bibfnamefont {S.}~\bibnamefont {Galtier}},
  \bibinfo {author} {\bibfnamefont {S.}~\bibnamefont {Thomas}}, \bibinfo
  {author} {\bibfnamefont {M.}~\bibnamefont {Bonnaud}}, \bibinfo {author}
  {\bibfnamefont {L.}~\bibnamefont {Julien}}, \bibinfo {author} {\bibfnamefont
  {F.~m.~c.}\ \bibnamefont {Biraben}}, \bibinfo {author} {\bibfnamefont
  {F.~m.~c.}\ \bibnamefont {Nez}}, \bibinfo {author} {\bibfnamefont
  {M.}~\bibnamefont {Abgrall}},\ and\ \bibinfo {author} {\bibfnamefont
  {J.}~\bibnamefont {Gu\'ena}},\ }\href
  {https://doi.org/10.1103/PhysRevLett.120.183001} {\bibfield  {journal}
  {\bibinfo  {journal} {Physical Review Letters}\ }\textbf {\bibinfo {volume}
  {120}},\ \bibinfo {pages} {183001} (\bibinfo {year} {2018})}\BibitemShut
  {NoStop}%
\bibitem [{\citenamefont {Grinin}\ \emph {et~al.}(2020)\citenamefont {Grinin},
  \citenamefont {Matveev}, \citenamefont {Yost}, \citenamefont {Maisenbacher},
  \citenamefont {Wirthl}, \citenamefont {Pohl}, \citenamefont {Hänsch},\ and\
  \citenamefont {Udem}}]{Grinin2020}%
  \BibitemOpen
  \bibfield  {author} {\bibinfo {author} {\bibfnamefont {A.}~\bibnamefont
  {Grinin}}, \bibinfo {author} {\bibfnamefont {A.}~\bibnamefont {Matveev}},
  \bibinfo {author} {\bibfnamefont {D.~C.}\ \bibnamefont {Yost}}, \bibinfo
  {author} {\bibfnamefont {L.}~\bibnamefont {Maisenbacher}}, \bibinfo {author}
  {\bibfnamefont {V.}~\bibnamefont {Wirthl}}, \bibinfo {author} {\bibfnamefont
  {R.}~\bibnamefont {Pohl}}, \bibinfo {author} {\bibfnamefont {T.~W.}\
  \bibnamefont {Hänsch}},\ and\ \bibinfo {author} {\bibfnamefont
  {T.}~\bibnamefont {Udem}},\ }\href {https://doi.org/10.1126/science.abc7776}
  {\bibfield  {journal} {\bibinfo  {journal} {Science}\ }\textbf {\bibinfo
  {volume} {370}},\ \bibinfo {pages} {1061} (\bibinfo {year}
  {2020})}\BibitemShut {NoStop}%
\bibitem [{\citenamefont {Bezginov}\ \emph {et~al.}(2019)\citenamefont
  {Bezginov}, \citenamefont {Valdez}, \citenamefont {Horbatsch}, \citenamefont
  {Marsman}, \citenamefont {Vutha},\ and\ \citenamefont
  {Hessels}}]{Bezginov2019}%
  \BibitemOpen
  \bibfield  {author} {\bibinfo {author} {\bibfnamefont {N.}~\bibnamefont
  {Bezginov}}, \bibinfo {author} {\bibfnamefont {T.}~\bibnamefont {Valdez}},
  \bibinfo {author} {\bibfnamefont {M.}~\bibnamefont {Horbatsch}}, \bibinfo
  {author} {\bibfnamefont {A.}~\bibnamefont {Marsman}}, \bibinfo {author}
  {\bibfnamefont {A.~C.}\ \bibnamefont {Vutha}},\ and\ \bibinfo {author}
  {\bibfnamefont {E.~A.}\ \bibnamefont {Hessels}},\ }\href
  {https://doi.org/10.1126/science.aau7807} {\bibfield  {journal} {\bibinfo
  {journal} {Science}\ }\textbf {\bibinfo {volume} {365}},\ \bibinfo {pages}
  {1007} (\bibinfo {year} {2019})}\BibitemShut {NoStop}%
\bibitem [{\citenamefont {Beyer}\ \emph {et~al.}(2017)\citenamefont {Beyer},
  \citenamefont {Maisenbacher}, \citenamefont {Matveev}, \citenamefont {Pohl},
  \citenamefont {Khabarova}, \citenamefont {Grinin}, \citenamefont {Lamour},
  \citenamefont {Yost}, \citenamefont {Hänsch}, \citenamefont {Kolachevsky},\
  and\ \citenamefont {Udem}}]{Beyer2017}%
  \BibitemOpen
  \bibfield  {author} {\bibinfo {author} {\bibfnamefont {A.}~\bibnamefont
  {Beyer}}, \bibinfo {author} {\bibfnamefont {L.}~\bibnamefont {Maisenbacher}},
  \bibinfo {author} {\bibfnamefont {A.}~\bibnamefont {Matveev}}, \bibinfo
  {author} {\bibfnamefont {R.}~\bibnamefont {Pohl}}, \bibinfo {author}
  {\bibfnamefont {K.}~\bibnamefont {Khabarova}}, \bibinfo {author}
  {\bibfnamefont {A.}~\bibnamefont {Grinin}}, \bibinfo {author} {\bibfnamefont
  {T.}~\bibnamefont {Lamour}}, \bibinfo {author} {\bibfnamefont {D.~C.}\
  \bibnamefont {Yost}}, \bibinfo {author} {\bibfnamefont {T.~W.}\ \bibnamefont
  {Hänsch}}, \bibinfo {author} {\bibfnamefont {N.}~\bibnamefont
  {Kolachevsky}},\ and\ \bibinfo {author} {\bibfnamefont {T.}~\bibnamefont
  {Udem}},\ }\href {https://doi.org/10.1126/science.aah6677} {\bibfield
  {journal} {\bibinfo  {journal} {Science}\ }\textbf {\bibinfo {volume}
  {358}},\ \bibinfo {pages} {79} (\bibinfo {year} {2017})}\BibitemShut
  {NoStop}%
\bibitem [{\citenamefont {Schwob}\ \emph {et~al.}(1999)\citenamefont {Schwob},
  \citenamefont {Jozefowski}, \citenamefont {de~Beauvoir}, \citenamefont
  {Hilico}, \citenamefont {Nez}, \citenamefont {Julien}, \citenamefont
  {Biraben}, \citenamefont {Acef}, \citenamefont {Zondy},\ and\ \citenamefont
  {Clairon}}]{Schwob1999}%
  \BibitemOpen
  \bibfield  {author} {\bibinfo {author} {\bibfnamefont {C.}~\bibnamefont
  {Schwob}}, \bibinfo {author} {\bibfnamefont {L.}~\bibnamefont {Jozefowski}},
  \bibinfo {author} {\bibfnamefont {B.}~\bibnamefont {de~Beauvoir}}, \bibinfo
  {author} {\bibfnamefont {L.}~\bibnamefont {Hilico}}, \bibinfo {author}
  {\bibfnamefont {F.}~\bibnamefont {Nez}}, \bibinfo {author} {\bibfnamefont
  {L.}~\bibnamefont {Julien}}, \bibinfo {author} {\bibfnamefont
  {F.}~\bibnamefont {Biraben}}, \bibinfo {author} {\bibfnamefont
  {O.}~\bibnamefont {Acef}}, \bibinfo {author} {\bibfnamefont {J.-J.}\
  \bibnamefont {Zondy}},\ and\ \bibinfo {author} {\bibfnamefont
  {A.}~\bibnamefont {Clairon}},\ }\href
  {https://doi.org/10.1103/PhysRevLett.82.4960} {\bibfield  {journal} {\bibinfo
   {journal} {Physical Review Letters}\ }\textbf {\bibinfo {volume} {82}},\
  \bibinfo {pages} {4960} (\bibinfo {year} {1999})}\BibitemShut {NoStop}%
\bibitem [{\citenamefont {de~Beauvoir}\ \emph {et~al.}(1997)\citenamefont
  {de~Beauvoir}, \citenamefont {Nez}, \citenamefont {Julien}, \citenamefont
  {Cagnac}, \citenamefont {Biraben}, \citenamefont {Touahri}, \citenamefont
  {Hilico}, \citenamefont {Acef}, \citenamefont {Clairon},\ and\ \citenamefont
  {Zondy}}]{Beauvior1997}%
  \BibitemOpen
  \bibfield  {author} {\bibinfo {author} {\bibfnamefont {B.}~\bibnamefont
  {de~Beauvoir}}, \bibinfo {author} {\bibfnamefont {F.}~\bibnamefont {Nez}},
  \bibinfo {author} {\bibfnamefont {L.}~\bibnamefont {Julien}}, \bibinfo
  {author} {\bibfnamefont {B.}~\bibnamefont {Cagnac}}, \bibinfo {author}
  {\bibfnamefont {F.}~\bibnamefont {Biraben}}, \bibinfo {author} {\bibfnamefont
  {D.}~\bibnamefont {Touahri}}, \bibinfo {author} {\bibfnamefont
  {L.}~\bibnamefont {Hilico}}, \bibinfo {author} {\bibfnamefont
  {O.}~\bibnamefont {Acef}}, \bibinfo {author} {\bibfnamefont {A.}~\bibnamefont
  {Clairon}},\ and\ \bibinfo {author} {\bibfnamefont {J.~J.}\ \bibnamefont
  {Zondy}},\ }\href {https://doi.org/10.1103/PhysRevLett.78.440} {\bibfield
  {journal} {\bibinfo  {journal} {Physical Review Letters}\ }\textbf {\bibinfo
  {volume} {78}},\ \bibinfo {pages} {440} (\bibinfo {year} {1997})}\BibitemShut
  {NoStop}%
\bibitem [{\citenamefont {de~Beauvoir}\ \emph {et~al.}(2000)\citenamefont
  {de~Beauvoir}, \citenamefont {Schwob}, \citenamefont {Acef}, \citenamefont
  {Jozefowski}, \citenamefont {Hilico}, \citenamefont {Nez}, \citenamefont
  {Julien}, \citenamefont {Calairon},\ and\ \citenamefont
  {Biraben}}]{deBeauvoir2000}%
  \BibitemOpen
  \bibfield  {author} {\bibinfo {author} {\bibfnamefont {B.}~\bibnamefont
  {de~Beauvoir}}, \bibinfo {author} {\bibfnamefont {C.}~\bibnamefont {Schwob}},
  \bibinfo {author} {\bibfnamefont {O.}~\bibnamefont {Acef}}, \bibinfo {author}
  {\bibfnamefont {L.}~\bibnamefont {Jozefowski}}, \bibinfo {author}
  {\bibfnamefont {L.}~\bibnamefont {Hilico}}, \bibinfo {author} {\bibfnamefont
  {F.}~\bibnamefont {Nez}}, \bibinfo {author} {\bibfnamefont {L.}~\bibnamefont
  {Julien}}, \bibinfo {author} {\bibfnamefont {A.}~\bibnamefont {Calairon}},\
  and\ \bibinfo {author} {\bibfnamefont {F.}~\bibnamefont {Biraben}},\ }\href
  {https://doi.org/10.1007/s100530070043} {\bibfield  {journal} {\bibinfo
  {journal} {Eur. Phys. J. D}\ }\textbf {\bibinfo {volume} {12}},\ \bibinfo
  {pages} {1434} (\bibinfo {year} {2000})}\BibitemShut {NoStop}%
\bibitem [{\citenamefont {Derevianko}\ and\ \citenamefont
  {Katori}(2011)}]{Derevianko2011}%
  \BibitemOpen
  \bibfield  {author} {\bibinfo {author} {\bibfnamefont {A.}~\bibnamefont
  {Derevianko}}\ and\ \bibinfo {author} {\bibfnamefont {H.}~\bibnamefont
  {Katori}},\ }\href {https://doi.org/10.1103/RevModPhys.83.331} {\bibfield
  {journal} {\bibinfo  {journal} {Reviews of Modern Physics}\ }\textbf
  {\bibinfo {volume} {83}},\ \bibinfo {pages} {331} (\bibinfo {year}
  {2011})}\BibitemShut {NoStop}%
\bibitem [{\citenamefont {Takamoto}\ \emph {et~al.}(2005)\citenamefont
  {Takamoto}, \citenamefont {Hong}, \citenamefont {Higashi},\ and\
  \citenamefont {Katori}}]{Takamoto2005}%
  \BibitemOpen
  \bibfield  {author} {\bibinfo {author} {\bibfnamefont {M.}~\bibnamefont
  {Takamoto}}, \bibinfo {author} {\bibfnamefont {F.-L.}\ \bibnamefont {Hong}},
  \bibinfo {author} {\bibfnamefont {R.}~\bibnamefont {Higashi}},\ and\ \bibinfo
  {author} {\bibfnamefont {H.}~\bibnamefont {Katori}},\ }\href
  {https://doi.org/10.1038/nature03541} {\bibfield  {journal} {\bibinfo
  {journal} {Nature}\ }\textbf {\bibinfo {volume} {435}},\ \bibinfo {pages}
  {321} (\bibinfo {year} {2005})}\BibitemShut {NoStop}%
\bibitem [{\citenamefont {Le~Targat}\ \emph {et~al.}(2006)\citenamefont
  {Le~Targat}, \citenamefont {Baillard}, \citenamefont {Fouch{\'e}},
  \citenamefont {Brusch}, \citenamefont {Tcherbakoff}, \citenamefont {Rovera},\
  and\ \citenamefont {Lemonde}}]{Le2006}%
  \BibitemOpen
  \bibfield  {author} {\bibinfo {author} {\bibfnamefont {R.}~\bibnamefont
  {Le~Targat}}, \bibinfo {author} {\bibfnamefont {X.}~\bibnamefont {Baillard}},
  \bibinfo {author} {\bibfnamefont {M.}~\bibnamefont {Fouch{\'e}}}, \bibinfo
  {author} {\bibfnamefont {A.}~\bibnamefont {Brusch}}, \bibinfo {author}
  {\bibfnamefont {O.}~\bibnamefont {Tcherbakoff}}, \bibinfo {author}
  {\bibfnamefont {G.~D.}\ \bibnamefont {Rovera}},\ and\ \bibinfo {author}
  {\bibfnamefont {P.}~\bibnamefont {Lemonde}},\ }\href
  {https://doi.org/10.1103/PhysRevLett.97.130801} {\bibfield  {journal}
  {\bibinfo  {journal} {Physical Review Letters}\ }\textbf {\bibinfo {volume}
  {97}},\ \bibinfo {pages} {130801} (\bibinfo {year} {2006})}\BibitemShut
  {NoStop}%
\bibitem [{\citenamefont {Ludlow}\ \emph {et~al.}(2006)\citenamefont {Ludlow},
  \citenamefont {Boyd}, \citenamefont {Zelevinsky}, \citenamefont {Foreman},
  \citenamefont {Blatt}, \citenamefont {Notcutt}, \citenamefont {Ido},\ and\
  \citenamefont {Ye}}]{Ludlow2006}%
  \BibitemOpen
  \bibfield  {author} {\bibinfo {author} {\bibfnamefont {A.~D.}\ \bibnamefont
  {Ludlow}}, \bibinfo {author} {\bibfnamefont {M.~M.}\ \bibnamefont {Boyd}},
  \bibinfo {author} {\bibfnamefont {T.}~\bibnamefont {Zelevinsky}}, \bibinfo
  {author} {\bibfnamefont {S.~M.}\ \bibnamefont {Foreman}}, \bibinfo {author}
  {\bibfnamefont {S.}~\bibnamefont {Blatt}}, \bibinfo {author} {\bibfnamefont
  {M.}~\bibnamefont {Notcutt}}, \bibinfo {author} {\bibfnamefont
  {T.}~\bibnamefont {Ido}},\ and\ \bibinfo {author} {\bibfnamefont
  {J.}~\bibnamefont {Ye}},\ }\href
  {https://doi.org/10.1103/PhysRevLett.96.033003} {\bibfield  {journal}
  {\bibinfo  {journal} {Physical Review Letters}\ }\textbf {\bibinfo {volume}
  {96}},\ \bibinfo {pages} {033003} (\bibinfo {year} {2006})}\BibitemShut
  {NoStop}%
\bibitem [{\citenamefont {Barber}\ \emph {et~al.}(2006)\citenamefont {Barber},
  \citenamefont {Hoyt}, \citenamefont {Oates}, \citenamefont {Hollberg},
  \citenamefont {Taichenachev},\ and\ \citenamefont {Yudin}}]{Barber2006}%
  \BibitemOpen
  \bibfield  {author} {\bibinfo {author} {\bibfnamefont {Z.~W.}\ \bibnamefont
  {Barber}}, \bibinfo {author} {\bibfnamefont {C.~W.}\ \bibnamefont {Hoyt}},
  \bibinfo {author} {\bibfnamefont {C.~W.}\ \bibnamefont {Oates}}, \bibinfo
  {author} {\bibfnamefont {L.}~\bibnamefont {Hollberg}}, \bibinfo {author}
  {\bibfnamefont {A.~V.}\ \bibnamefont {Taichenachev}},\ and\ \bibinfo {author}
  {\bibfnamefont {V.~I.}\ \bibnamefont {Yudin}},\ }\href
  {https://doi.org/10.1103/PhysRevLett.96.083002} {\bibfield  {journal}
  {\bibinfo  {journal} {Physical Review Letters}\ }\textbf {\bibinfo {volume}
  {96}},\ \bibinfo {pages} {083002} (\bibinfo {year} {2006})}\BibitemShut
  {NoStop}%
\bibitem [{\citenamefont {Bothwell}\ \emph {et~al.}(2019)\citenamefont
  {Bothwell}, \citenamefont {Kedar}, \citenamefont {Oelker}, \citenamefont
  {Robinson}, \citenamefont {Bromley}, \citenamefont {Tew}, \citenamefont
  {Ye},\ and\ \citenamefont {Kennedy}}]{Bothwell2019}%
  \BibitemOpen
  \bibfield  {author} {\bibinfo {author} {\bibfnamefont {T.}~\bibnamefont
  {Bothwell}}, \bibinfo {author} {\bibfnamefont {D.}~\bibnamefont {Kedar}},
  \bibinfo {author} {\bibfnamefont {E.}~\bibnamefont {Oelker}}, \bibinfo
  {author} {\bibfnamefont {J.~M.}\ \bibnamefont {Robinson}}, \bibinfo {author}
  {\bibfnamefont {S.~L.}\ \bibnamefont {Bromley}}, \bibinfo {author}
  {\bibfnamefont {W.~L.}\ \bibnamefont {Tew}}, \bibinfo {author} {\bibfnamefont
  {J.}~\bibnamefont {Ye}},\ and\ \bibinfo {author} {\bibfnamefont {C.~J.}\
  \bibnamefont {Kennedy}},\ }\href {https://doi.org/10.1088/1681-7575/ab4089}
  {\bibfield  {journal} {\bibinfo  {journal} {Metrologia}\ }\textbf {\bibinfo
  {volume} {56}},\ \bibinfo {pages} {065004} (\bibinfo {year}
  {2019})}\BibitemShut {NoStop}%
\bibitem [{\citenamefont {Crivelli}\ and\ \citenamefont
  {Kolachevsky}(2020)}]{Crivelli2020}%
  \BibitemOpen
  \bibfield  {author} {\bibinfo {author} {\bibfnamefont {P.}~\bibnamefont
  {Crivelli}}\ and\ \bibinfo {author} {\bibfnamefont {N.}~\bibnamefont
  {Kolachevsky}},\ }\href {https://doi.org/10.1007/s10751-018-1549-4}
  {\bibfield  {journal} {\bibinfo  {journal} {Hyperfine Interactions}\ }\textbf
  {\bibinfo {volume} {241}},\ \bibinfo {pages} {1} (\bibinfo {year}
  {2020})}\BibitemShut {NoStop}%
\bibitem [{\citenamefont {Adhikari}\ \emph {et~al.}(2016)\citenamefont
  {Adhikari}, \citenamefont {Kawasaki},\ and\ \citenamefont
  {Jentschura}}]{Adhikari2016}%
  \BibitemOpen
  \bibfield  {author} {\bibinfo {author} {\bibfnamefont {C.~M.}\ \bibnamefont
  {Adhikari}}, \bibinfo {author} {\bibfnamefont {A.}~\bibnamefont {Kawasaki}},\
  and\ \bibinfo {author} {\bibfnamefont {U.~D.}\ \bibnamefont {Jentschura}},\
  }\href {https://doi.org/10.1103/PhysRevA.94.032510} {\bibfield  {journal}
  {\bibinfo  {journal} {Physical Review A}\ }\textbf {\bibinfo {volume} {94}},\
  \bibinfo {pages} {032510} (\bibinfo {year} {2016})}\BibitemShut {NoStop}%
\bibitem [{\citenamefont {Adhikari}\ \emph {et~al.}(2022)\citenamefont
  {Adhikari}, \citenamefont {Canales}, \citenamefont {Arthanayaka},\ and\
  \citenamefont {Jentschura}}]{Adhikari2022}%
  \BibitemOpen
  \bibfield  {author} {\bibinfo {author} {\bibfnamefont {C.~M.}\ \bibnamefont
  {Adhikari}}, \bibinfo {author} {\bibfnamefont {J.~C.}\ \bibnamefont
  {Canales}}, \bibinfo {author} {\bibfnamefont {T.~P.}\ \bibnamefont
  {Arthanayaka}},\ and\ \bibinfo {author} {\bibfnamefont {U.~D.}\ \bibnamefont
  {Jentschura}},\ }\href {https://doi.org/10.3390/atoms10010001} {\bibfield
  {journal} {\bibinfo  {journal} {Atoms}\ }\textbf {\bibinfo {volume} {10}},\
  \bibinfo {pages} {1} (\bibinfo {year} {2022})}\BibitemShut {NoStop}%
\bibitem [{\citenamefont {Holt}\ and\ \citenamefont {Sellin}(1972)}]{Holt1972}%
  \BibitemOpen
  \bibfield  {author} {\bibinfo {author} {\bibfnamefont {H.~K.}\ \bibnamefont
  {Holt}}\ and\ \bibinfo {author} {\bibfnamefont {I.}~\bibnamefont {Sellin}},\
  }\href@noop {} {\bibfield  {journal} {\bibinfo  {journal} {Physical Review
  A}\ }\textbf {\bibinfo {volume} {6}},\ \bibinfo {pages} {508} (\bibinfo
  {year} {1972})}\BibitemShut {NoStop}%
\bibitem [{\citenamefont {D{\"o}rscher}\ \emph {et~al.}(2018)\citenamefont
  {D{\"o}rscher}, \citenamefont {Schwarz}, \citenamefont {Al-Masoudi},
  \citenamefont {Falke}, \citenamefont {Sterr},\ and\ \citenamefont
  {Lisdat}}]{Dorscher2018}%
  \BibitemOpen
  \bibfield  {author} {\bibinfo {author} {\bibfnamefont {S.}~\bibnamefont
  {D{\"o}rscher}}, \bibinfo {author} {\bibfnamefont {R.}~\bibnamefont
  {Schwarz}}, \bibinfo {author} {\bibfnamefont {A.}~\bibnamefont {Al-Masoudi}},
  \bibinfo {author} {\bibfnamefont {S.}~\bibnamefont {Falke}}, \bibinfo
  {author} {\bibfnamefont {U.}~\bibnamefont {Sterr}},\ and\ \bibinfo {author}
  {\bibfnamefont {C.}~\bibnamefont {Lisdat}},\ }\href
  {https://doi.org/10.1103/PhysRevA.97.063419} {\bibfield  {journal} {\bibinfo
  {journal} {Physical Review A}\ }\textbf {\bibinfo {volume} {97}},\ \bibinfo
  {pages} {063419} (\bibinfo {year} {2018})}\BibitemShut {NoStop}%
\bibitem [{\citenamefont {Zernik}(1964{\natexlab{a}})}]{Zernik1964}%
  \BibitemOpen
  \bibfield  {author} {\bibinfo {author} {\bibfnamefont {W.}~\bibnamefont
  {Zernik}},\ }\href@noop {} {\bibfield  {journal} {\bibinfo  {journal}
  {Physical Review}\ }\textbf {\bibinfo {volume} {133}},\ \bibinfo {pages}
  {A117} (\bibinfo {year} {1964}{\natexlab{a}})}\BibitemShut {NoStop}%
\bibitem [{\citenamefont {Gontier}\ and\ \citenamefont
  {Trahin}(1971)}]{Gontier1971}%
  \BibitemOpen
  \bibfield  {author} {\bibinfo {author} {\bibfnamefont {Y.}~\bibnamefont
  {Gontier}}\ and\ \bibinfo {author} {\bibfnamefont {M.}~\bibnamefont
  {Trahin}},\ }\href@noop {} {\bibfield  {journal} {\bibinfo  {journal}
  {Physical Review A}\ }\textbf {\bibinfo {volume} {4}},\ \bibinfo {pages}
  {1896} (\bibinfo {year} {1971})}\BibitemShut {NoStop}%
\bibitem [{\citenamefont {Klarsfeld}(1972)}]{Klarsfeld1972}%
  \BibitemOpen
  \bibfield  {author} {\bibinfo {author} {\bibfnamefont {S.}~\bibnamefont
  {Klarsfeld}},\ }\href@noop {} {\bibfield  {journal} {\bibinfo  {journal}
  {Physical Review A}\ }\textbf {\bibinfo {volume} {6}},\ \bibinfo {pages}
  {506} (\bibinfo {year} {1972})}\BibitemShut {NoStop}%
\bibitem [{\citenamefont {Bachau}\ \emph {et~al.}(2017)\citenamefont {Bachau},
  \citenamefont {Dondera}, \citenamefont {Florescu},\ and\ \citenamefont
  {Marian}}]{Bachau2017}%
  \BibitemOpen
  \bibfield  {author} {\bibinfo {author} {\bibfnamefont {H.}~\bibnamefont
  {Bachau}}, \bibinfo {author} {\bibfnamefont {M.}~\bibnamefont {Dondera}},
  \bibinfo {author} {\bibfnamefont {V.}~\bibnamefont {Florescu}},\ and\
  \bibinfo {author} {\bibfnamefont {T.~A.}\ \bibnamefont {Marian}},\
  }\href@noop {} {\bibfield  {journal} {\bibinfo  {journal} {Journal of Physics
  B: Atomic, Molecular and Optical Physics}\ }\textbf {\bibinfo {volume}
  {50}},\ \bibinfo {pages} {174003} (\bibinfo {year} {2017})}\BibitemShut
  {NoStop}%
\bibitem [{\citenamefont {Heno}\ \emph {et~al.}(1980)\citenamefont {Heno},
  \citenamefont {Maquet},\ and\ \citenamefont {Schwarcz}}]{Heno1980}%
  \BibitemOpen
  \bibfield  {author} {\bibinfo {author} {\bibfnamefont {Y.}~\bibnamefont
  {Heno}}, \bibinfo {author} {\bibfnamefont {A.}~\bibnamefont {Maquet}},\ and\
  \bibinfo {author} {\bibfnamefont {R.}~\bibnamefont {Schwarcz}},\ }\href@noop
  {} {\bibfield  {journal} {\bibinfo  {journal} {Journal of Applied Physics}\
  }\textbf {\bibinfo {volume} {51}},\ \bibinfo {pages} {11} (\bibinfo {year}
  {1980})}\BibitemShut {NoStop}%
\bibitem [{\citenamefont {Zernik}(1964{\natexlab{b}})}]{zernik1964two}%
  \BibitemOpen
  \bibfield  {author} {\bibinfo {author} {\bibfnamefont {W.}~\bibnamefont
  {Zernik}},\ }\href@noop {} {\bibfield  {journal} {\bibinfo  {journal}
  {Physical Review}\ }\textbf {\bibinfo {volume} {135}},\ \bibinfo {pages}
  {A51} (\bibinfo {year} {1964}{\natexlab{b}})}\BibitemShut {NoStop}%
\bibitem [{\citenamefont {Rapoport}\ \emph {et~al.}(1969)\citenamefont
  {Rapoport}, \citenamefont {Zon},\ and\ \citenamefont
  {Manakov}}]{rapoport1969}%
  \BibitemOpen
  \bibfield  {author} {\bibinfo {author} {\bibfnamefont {L.}~\bibnamefont
  {Rapoport}}, \bibinfo {author} {\bibfnamefont {B.}~\bibnamefont {Zon}},\ and\
  \bibinfo {author} {\bibfnamefont {L.}~\bibnamefont {Manakov}},\ }\href@noop
  {} {\bibfield  {journal} {\bibinfo  {journal} {Sov. Phys. JETP}\ }\textbf
  {\bibinfo {volume} {29}},\ \bibinfo {pages} {220} (\bibinfo {year}
  {1969})}\BibitemShut {NoStop}%
\bibitem [{\citenamefont {Khristenko}\ and\ \citenamefont
  {Vetchinkin}(1976)}]{khristenko1976}%
  \BibitemOpen
  \bibfield  {author} {\bibinfo {author} {\bibfnamefont {S.}~\bibnamefont
  {Khristenko}}\ and\ \bibinfo {author} {\bibfnamefont {S.}~\bibnamefont
  {Vetchinkin}},\ }\href@noop {} {\bibfield  {journal} {\bibinfo  {journal}
  {Opt. Spectrosc.(USSR)(Engl. Transl.);(United States)}\ }\textbf {\bibinfo
  {volume} {40}} (\bibinfo {year} {1976})}\BibitemShut {NoStop}%
\bibitem [{\citenamefont {Kassaee}\ \emph {et~al.}(1988)\citenamefont
  {Kassaee}, \citenamefont {Rustgi},\ and\ \citenamefont {Long}}]{kassaee1988}%
  \BibitemOpen
  \bibfield  {author} {\bibinfo {author} {\bibfnamefont {A.}~\bibnamefont
  {Kassaee}}, \bibinfo {author} {\bibfnamefont {M.}~\bibnamefont {Rustgi}},\
  and\ \bibinfo {author} {\bibfnamefont {S.}~\bibnamefont {Long}},\ }\href@noop
  {} {\bibfield  {journal} {\bibinfo  {journal} {Physical Review A}\ }\textbf
  {\bibinfo {volume} {37}},\ \bibinfo {pages} {999} (\bibinfo {year}
  {1988})}\BibitemShut {NoStop}%
\bibitem [{\citenamefont {Karule}\ and\ \citenamefont
  {Moine}(2003)}]{karule2003}%
  \BibitemOpen
  \bibfield  {author} {\bibinfo {author} {\bibfnamefont {E.}~\bibnamefont
  {Karule}}\ and\ \bibinfo {author} {\bibfnamefont {B.}~\bibnamefont {Moine}},\
  }\href@noop {} {\bibfield  {journal} {\bibinfo  {journal} {Journal of Physics
  B: Atomic, Molecular and Optical Physics}\ }\textbf {\bibinfo {volume}
  {36}},\ \bibinfo {pages} {1963} (\bibinfo {year} {2003})}\BibitemShut
  {NoStop}%
\bibitem [{\citenamefont {Vázquez-Carson}\ \emph {et~al.}(2022)\citenamefont
  {Vázquez-Carson}, \citenamefont {Sun}, \citenamefont {Dai}, \citenamefont
  {Mitra},\ and\ \citenamefont {Zelevinsky}}]{VazquezCarson2022}%
  \BibitemOpen
  \bibfield  {author} {\bibinfo {author} {\bibfnamefont {S.~F.}\ \bibnamefont
  {Vázquez-Carson}}, \bibinfo {author} {\bibfnamefont {Q.}~\bibnamefont
  {Sun}}, \bibinfo {author} {\bibfnamefont {J.}~\bibnamefont {Dai}}, \bibinfo
  {author} {\bibfnamefont {D.}~\bibnamefont {Mitra}},\ and\ \bibinfo {author}
  {\bibfnamefont {T.}~\bibnamefont {Zelevinsky}},\ }\href
  {https://doi.org/10.1088/1367-2630/ac806c} {\bibfield  {journal} {\bibinfo
  {journal} {New Journal of Physics}\ }\textbf {\bibinfo {volume} {24}},\
  \bibinfo {pages} {083006} (\bibinfo {year} {2022})}\BibitemShut {NoStop}%
\bibitem [{\citenamefont {Baker}\ \emph {et~al.}(2021)\citenamefont {Baker},
  \citenamefont {Bertsche}, \citenamefont {Capra}, \citenamefont {Carruth},
  \citenamefont {Cesar}, \citenamefont {Charlton}, \citenamefont {Christensen},
  \citenamefont {Collister}, \citenamefont {Mathad}, \citenamefont {Eriksson}
  \emph {et~al.}}]{Baker2021}%
  \BibitemOpen
  \bibfield  {author} {\bibinfo {author} {\bibfnamefont {C.}~\bibnamefont
  {Baker}}, \bibinfo {author} {\bibfnamefont {W.}~\bibnamefont {Bertsche}},
  \bibinfo {author} {\bibfnamefont {A.}~\bibnamefont {Capra}}, \bibinfo
  {author} {\bibfnamefont {C.}~\bibnamefont {Carruth}}, \bibinfo {author}
  {\bibfnamefont {C.}~\bibnamefont {Cesar}}, \bibinfo {author} {\bibfnamefont
  {M.}~\bibnamefont {Charlton}}, \bibinfo {author} {\bibfnamefont
  {A.}~\bibnamefont {Christensen}}, \bibinfo {author} {\bibfnamefont
  {R.}~\bibnamefont {Collister}}, \bibinfo {author} {\bibfnamefont {A.~C.}\
  \bibnamefont {Mathad}}, \bibinfo {author} {\bibfnamefont {S.}~\bibnamefont
  {Eriksson}}, \emph {et~al.},\ }\href
  {https://doi.org/10.1038/s41586-021-03289-6} {\bibfield  {journal} {\bibinfo
  {journal} {Nature}\ }\textbf {\bibinfo {volume} {592}},\ \bibinfo {pages}
  {35} (\bibinfo {year} {2021})}\BibitemShut {NoStop}%
\bibitem [{\citenamefont {Lane}(2015)}]{Lane2015}%
  \BibitemOpen
  \bibfield  {author} {\bibinfo {author} {\bibfnamefont {I.~C.}\ \bibnamefont
  {Lane}},\ }\href {https://doi.org/10.1103/PhysRevA.92.022511} {\bibfield
  {journal} {\bibinfo  {journal} {Physical Review A}\ }\textbf {\bibinfo
  {volume} {92}},\ \bibinfo {pages} {022511} (\bibinfo {year}
  {2015})}\BibitemShut {NoStop}%
\bibitem [{\citenamefont {Gabrielse}\ \emph {et~al.}(2018)\citenamefont
  {Gabrielse}, \citenamefont {Glowacz}, \citenamefont {Grzonka}, \citenamefont
  {Hamley}, \citenamefont {Hessels}, \citenamefont {Jones}, \citenamefont
  {Khatri}, \citenamefont {Lee}, \citenamefont {Meisenhelder}, \citenamefont
  {Morrison} \emph {et~al.}}]{Gabrielse2018}%
  \BibitemOpen
  \bibfield  {author} {\bibinfo {author} {\bibfnamefont {G.}~\bibnamefont
  {Gabrielse}}, \bibinfo {author} {\bibfnamefont {B.}~\bibnamefont {Glowacz}},
  \bibinfo {author} {\bibfnamefont {D.}~\bibnamefont {Grzonka}}, \bibinfo
  {author} {\bibfnamefont {C.}~\bibnamefont {Hamley}}, \bibinfo {author}
  {\bibfnamefont {E.}~\bibnamefont {Hessels}}, \bibinfo {author} {\bibfnamefont
  {N.}~\bibnamefont {Jones}}, \bibinfo {author} {\bibfnamefont
  {G.}~\bibnamefont {Khatri}}, \bibinfo {author} {\bibfnamefont
  {S.}~\bibnamefont {Lee}}, \bibinfo {author} {\bibfnamefont {C.}~\bibnamefont
  {Meisenhelder}}, \bibinfo {author} {\bibfnamefont {T.}~\bibnamefont
  {Morrison}}, \emph {et~al.},\ }\href {https://doi.org/10.1364/OL.43.002905}
  {\bibfield  {journal} {\bibinfo  {journal} {Optics letters}\ }\textbf
  {\bibinfo {volume} {43}},\ \bibinfo {pages} {2905} (\bibinfo {year}
  {2018})}\BibitemShut {NoStop}%
\bibitem [{\citenamefont {Burkley}\ and\ \citenamefont
  {Yost}(2018)}]{Burkley2018}%
  \BibitemOpen
  \bibfield  {author} {\bibinfo {author} {\bibfnamefont {Z.}~\bibnamefont
  {Burkley}}\ and\ \bibinfo {author} {\bibfnamefont {D.}~\bibnamefont {Yost}},\
  }\href@noop {} {\bibfield  {journal} {\bibinfo  {journal} {Bulletin of the
  American Physical Society}\ }\textbf {\bibinfo {volume} {63}} (\bibinfo
  {year} {2018})}\BibitemShut {NoStop}%
\bibitem [{Dra(2006)}]{Drake2006}%
  \BibitemOpen
  \href@noop {} {\emph {\bibinfo {title} {Springer handbook of atomic,
  molecular, and optical physics}}},\ \bibinfo {edition} {2nd}\ ed.\ (\bibinfo
  {publisher} {Springer},\ \bibinfo {year} {2006})\BibitemShut {NoStop}%
\bibitem [{\citenamefont {Grimm}\ \emph {et~al.}(2000)\citenamefont {Grimm},
  \citenamefont {Weidem{\"u}ller},\ and\ \citenamefont
  {Ovchinnikov}}]{Grimm2000}%
  \BibitemOpen
  \bibfield  {author} {\bibinfo {author} {\bibfnamefont {R.}~\bibnamefont
  {Grimm}}, \bibinfo {author} {\bibfnamefont {M.}~\bibnamefont
  {Weidem{\"u}ller}},\ and\ \bibinfo {author} {\bibfnamefont {Y.~B.}\
  \bibnamefont {Ovchinnikov}},\ }in\ \href
  {https://doi.org/10.1016/S1049-250X(08)60186-X} {\emph {\bibinfo {booktitle}
  {Advances in atomic, molecular, and optical physics}}},\ Vol.~\bibinfo
  {volume} {42}\ (\bibinfo  {publisher} {Elsevier},\ \bibinfo {year} {2000})\
  pp.\ \bibinfo {pages} {95--170}\BibitemShut {NoStop}%
\bibitem [{\citenamefont {Haas}\ \emph {et~al.}(2006)\citenamefont {Haas},
  \citenamefont {Jentschura},\ and\ \citenamefont {Keitel}}]{Haas2006}%
  \BibitemOpen
  \bibfield  {author} {\bibinfo {author} {\bibfnamefont {M.}~\bibnamefont
  {Haas}}, \bibinfo {author} {\bibfnamefont {U.~D.}\ \bibnamefont
  {Jentschura}},\ and\ \bibinfo {author} {\bibfnamefont {C.~H.}\ \bibnamefont
  {Keitel}},\ }\href {https://doi.org/10.1119/1.2140742} {\bibfield  {journal}
  {\bibinfo  {journal} {American Journal of Physics}\ }\textbf {\bibinfo
  {volume} {74}},\ \bibinfo {pages} {77} (\bibinfo {year} {2006})}\BibitemShut
  {NoStop}%
\bibitem [{\citenamefont {Le~Kien}\ \emph {et~al.}(2013)\citenamefont
  {Le~Kien}, \citenamefont {Schneeweiss},\ and\ \citenamefont
  {Rauschenbeutel}}]{LeKien2013}%
  \BibitemOpen
  \bibfield  {author} {\bibinfo {author} {\bibfnamefont {F.}~\bibnamefont
  {Le~Kien}}, \bibinfo {author} {\bibfnamefont {P.}~\bibnamefont
  {Schneeweiss}},\ and\ \bibinfo {author} {\bibfnamefont {A.}~\bibnamefont
  {Rauschenbeutel}},\ }\href {https://doi.org/10.1140/epjd/e2013-30729-x}
  {\bibfield  {journal} {\bibinfo  {journal} {The European Physical Journal D}\
  }\textbf {\bibinfo {volume} {67}},\ \bibinfo {pages} {1} (\bibinfo {year}
  {2013})}\BibitemShut {NoStop}%
\bibitem [{\citenamefont {Alnis}\ \emph {et~al.}(2008)\citenamefont {Alnis},
  \citenamefont {Matveev}, \citenamefont {Kolachevsky}, \citenamefont {Udem},\
  and\ \citenamefont {H\"ansch}}]{Alnis2008}%
  \BibitemOpen
  \bibfield  {author} {\bibinfo {author} {\bibfnamefont {J.}~\bibnamefont
  {Alnis}}, \bibinfo {author} {\bibfnamefont {A.}~\bibnamefont {Matveev}},
  \bibinfo {author} {\bibfnamefont {N.}~\bibnamefont {Kolachevsky}}, \bibinfo
  {author} {\bibfnamefont {T.}~\bibnamefont {Udem}},\ and\ \bibinfo {author}
  {\bibfnamefont {T.~W.}\ \bibnamefont {H\"ansch}},\ }\href
  {https://doi.org/10.1103/PhysRevA.77.053809} {\bibfield  {journal} {\bibinfo
  {journal} {Physical Review A}\ }\textbf {\bibinfo {volume} {77}},\ \bibinfo
  {pages} {053809} (\bibinfo {year} {2008})}\BibitemShut {NoStop}%
\bibitem [{\citenamefont {Br\"aunlich}\ and\ \citenamefont
  {Lambropoulos}(1970)}]{Braunlich1970}%
  \BibitemOpen
  \bibfield  {author} {\bibinfo {author} {\bibfnamefont {P.}~\bibnamefont
  {Br\"aunlich}}\ and\ \bibinfo {author} {\bibfnamefont {P.}~\bibnamefont
  {Lambropoulos}},\ }\href {https://doi.org/10.1103/PhysRevLett.25.135}
  {\bibfield  {journal} {\bibinfo  {journal} {Physical Review Letters}\
  }\textbf {\bibinfo {volume} {25}},\ \bibinfo {pages} {135} (\bibinfo {year}
  {1970})}\BibitemShut {NoStop}%
\bibitem [{\citenamefont {Potvliege}(1998)}]{Potvliege1998}%
  \BibitemOpen
  \bibfield  {author} {\bibinfo {author} {\bibfnamefont {R.}~\bibnamefont
  {Potvliege}},\ }\href@noop {} {\bibfield  {journal} {\bibinfo  {journal}
  {Computer physics communications}\ }\textbf {\bibinfo {volume} {114}},\
  \bibinfo {pages} {42} (\bibinfo {year} {1998})}\BibitemShut {NoStop}%
\bibitem [{\citenamefont {H{\"a}nsch}\ \emph {et~al.}(1975)\citenamefont
  {H{\"a}nsch}, \citenamefont {Lee}, \citenamefont {Wallenstein},\ and\
  \citenamefont {Wieman}}]{hansch1975}%
  \BibitemOpen
  \bibfield  {author} {\bibinfo {author} {\bibfnamefont {T.~W.}\ \bibnamefont
  {H{\"a}nsch}}, \bibinfo {author} {\bibfnamefont {S.~A.}\ \bibnamefont {Lee}},
  \bibinfo {author} {\bibfnamefont {R.}~\bibnamefont {Wallenstein}},\ and\
  \bibinfo {author} {\bibfnamefont {C.}~\bibnamefont {Wieman}},\ }\href@noop {}
  {\bibfield  {journal} {\bibinfo  {journal} {Physical Review Letters}\
  }\textbf {\bibinfo {volume} {34}},\ \bibinfo {pages} {307} (\bibinfo {year}
  {1975})}\BibitemShut {NoStop}%
\bibitem [{\citenamefont {Biraben}\ \emph {et~al.}(1991)\citenamefont
  {Biraben}, \citenamefont {Julien}, \citenamefont {Plon},\ and\ \citenamefont
  {Nez}}]{biraben1991}%
  \BibitemOpen
  \bibfield  {author} {\bibinfo {author} {\bibfnamefont {F.}~\bibnamefont
  {Biraben}}, \bibinfo {author} {\bibfnamefont {L.}~\bibnamefont {Julien}},
  \bibinfo {author} {\bibfnamefont {J.}~\bibnamefont {Plon}},\ and\ \bibinfo
  {author} {\bibfnamefont {F.}~\bibnamefont {Nez}},\ }\href@noop {} {\bibfield
  {journal} {\bibinfo  {journal} {Europhysics Letters}\ }\textbf {\bibinfo
  {volume} {15}},\ \bibinfo {pages} {831} (\bibinfo {year} {1991})}\BibitemShut
  {NoStop}%
\bibitem [{\citenamefont {Mart{\'\i}nez-Lahuerta}\ \emph
  {et~al.}(2022)\citenamefont {Mart{\'\i}nez-Lahuerta}, \citenamefont {Eilers},
  \citenamefont {Mehlst{\"a}ubler}, \citenamefont {Schmidt},\ and\
  \citenamefont {Hammerer}}]{martinez2022}%
  \BibitemOpen
  \bibfield  {author} {\bibinfo {author} {\bibfnamefont {V.}~\bibnamefont
  {Mart{\'\i}nez-Lahuerta}}, \bibinfo {author} {\bibfnamefont {S.}~\bibnamefont
  {Eilers}}, \bibinfo {author} {\bibfnamefont {T.}~\bibnamefont
  {Mehlst{\"a}ubler}}, \bibinfo {author} {\bibfnamefont {P.}~\bibnamefont
  {Schmidt}},\ and\ \bibinfo {author} {\bibfnamefont {K.}~\bibnamefont
  {Hammerer}},\ }\href@noop {} {\bibfield  {journal} {\bibinfo  {journal}
  {Physical Review A}\ }\textbf {\bibinfo {volume} {106}},\ \bibinfo {pages}
  {032803} (\bibinfo {year} {2022})}\BibitemShut {NoStop}%
\bibitem [{\citenamefont {Jitrik}\ and\ \citenamefont
  {Bunge}(2004)}]{jitrik2004transition}%
  \BibitemOpen
  \bibfield  {author} {\bibinfo {author} {\bibfnamefont {O.}~\bibnamefont
  {Jitrik}}\ and\ \bibinfo {author} {\bibfnamefont {C.~F.}\ \bibnamefont
  {Bunge}},\ }\href@noop {} {\bibfield  {journal} {\bibinfo  {journal} {Journal
  of physical and chemical reference data}\ }\textbf {\bibinfo {volume} {33}},\
  \bibinfo {pages} {1059} (\bibinfo {year} {2004})}\BibitemShut {NoStop}%
\bibitem [{\citenamefont {Seiler}\ \emph {et~al.}(2016)\citenamefont {Seiler},
  \citenamefont {Agner}, \citenamefont {Pillet},\ and\ \citenamefont
  {Merkt}}]{seiler2016}%
  \BibitemOpen
  \bibfield  {author} {\bibinfo {author} {\bibfnamefont {C.}~\bibnamefont
  {Seiler}}, \bibinfo {author} {\bibfnamefont {J.~A.}\ \bibnamefont {Agner}},
  \bibinfo {author} {\bibfnamefont {P.}~\bibnamefont {Pillet}},\ and\ \bibinfo
  {author} {\bibfnamefont {F.}~\bibnamefont {Merkt}},\ }\href@noop {}
  {\bibfield  {journal} {\bibinfo  {journal} {Journal of Physics B: Atomic,
  Molecular and Optical Physics}\ }\textbf {\bibinfo {volume} {49}},\ \bibinfo
  {pages} {094006} (\bibinfo {year} {2016})}\BibitemShut {NoStop}%
\bibitem [{\citenamefont {Vliegen}\ \emph {et~al.}(2007)\citenamefont
  {Vliegen}, \citenamefont {Hogan}, \citenamefont {Schmutz},\ and\
  \citenamefont {Merkt}}]{Vliegen2007}%
  \BibitemOpen
  \bibfield  {author} {\bibinfo {author} {\bibfnamefont {E.}~\bibnamefont
  {Vliegen}}, \bibinfo {author} {\bibfnamefont {S.~D.}\ \bibnamefont {Hogan}},
  \bibinfo {author} {\bibfnamefont {H.}~\bibnamefont {Schmutz}},\ and\ \bibinfo
  {author} {\bibfnamefont {F.}~\bibnamefont {Merkt}},\ }\href
  {https://doi.org/10.1103/PhysRevA.76.023405} {\bibfield  {journal} {\bibinfo
  {journal} {Physical Review A}\ }\textbf {\bibinfo {volume} {76}},\ \bibinfo
  {pages} {023405} (\bibinfo {year} {2007})}\BibitemShut {NoStop}%
\bibitem [{\citenamefont {Sa\ss{}mannshausen}\ \emph
  {et~al.}(2013)\citenamefont {Sa\ss{}mannshausen}, \citenamefont {Merkt},\
  and\ \citenamefont {Deiglmayr}}]{Sassmannhausen2013}%
  \BibitemOpen
  \bibfield  {author} {\bibinfo {author} {\bibfnamefont {H.}~\bibnamefont
  {Sa\ss{}mannshausen}}, \bibinfo {author} {\bibfnamefont {F.}~\bibnamefont
  {Merkt}},\ and\ \bibinfo {author} {\bibfnamefont {J.}~\bibnamefont
  {Deiglmayr}},\ }\href {https://doi.org/10.1103/PhysRevA.87.032519} {\bibfield
   {journal} {\bibinfo  {journal} {Physical Review A}\ }\textbf {\bibinfo
  {volume} {87}},\ \bibinfo {pages} {032519} (\bibinfo {year}
  {2013})}\BibitemShut {NoStop}%
\bibitem [{\citenamefont {Loudon}(2010)}]{Loudon2010}%
  \BibitemOpen
  \bibfield  {author} {\bibinfo {author} {\bibfnamefont {R.}~\bibnamefont
  {Loudon}},\ }\href@noop {} {\emph {\bibinfo {title} {The quantum theory of
  light}}},\ \bibinfo {edition} {3rd}\ ed.\ (\bibinfo  {publisher} {Oxford
  University Press},\ \bibinfo {year} {2010})\BibitemShut {NoStop}%
\bibitem [{\citenamefont {Potvliege}\ and\ \citenamefont
  {Shakeshaft}(1989)}]{Potvliege1989}%
  \BibitemOpen
  \bibfield  {author} {\bibinfo {author} {\bibfnamefont {R.}~\bibnamefont
  {Potvliege}}\ and\ \bibinfo {author} {\bibfnamefont {R.}~\bibnamefont
  {Shakeshaft}},\ }\href@noop {} {\bibfield  {journal} {\bibinfo  {journal}
  {Zeitschrift fuer Physik, D}\ }\textbf {\bibinfo {volume} {11}},\ \bibinfo
  {pages} {93} (\bibinfo {year} {1989})}\BibitemShut {NoStop}%
\bibitem [{\citenamefont {Mavromatis}(1991)}]{Mavromatis1991}%
  \BibitemOpen
  \bibfield  {author} {\bibinfo {author} {\bibfnamefont {H.~A.}\ \bibnamefont
  {Mavromatis}},\ }\href {https://doi.org/10.1119/1.16753} {\bibfield
  {journal} {\bibinfo  {journal} {American Journal of Physics}\ }\textbf
  {\bibinfo {volume} {59}},\ \bibinfo {pages} {738} (\bibinfo {year}
  {1991})}\BibitemShut {NoStop}%
\bibitem [{\citenamefont {Nandi}\ \emph {et~al.}(1996)\citenamefont {Nandi},
  \citenamefont {Bera}, \citenamefont {Panja},\ and\ \citenamefont
  {Talukdar}}]{Nandi1996}%
  \BibitemOpen
  \bibfield  {author} {\bibinfo {author} {\bibfnamefont {T.}~\bibnamefont
  {Nandi}}, \bibinfo {author} {\bibfnamefont {P.}~\bibnamefont {Bera}},
  \bibinfo {author} {\bibfnamefont {M.}~\bibnamefont {Panja}},\ and\ \bibinfo
  {author} {\bibfnamefont {B.}~\bibnamefont {Talukdar}},\ }\href
  {https://doi.org/10.1088/0305-4470/29/5/022} {\bibfield  {journal} {\bibinfo
  {journal} {Journal of Physics A: Mathematical and General}\ }\textbf
  {\bibinfo {volume} {29}},\ \bibinfo {pages} {1101} (\bibinfo {year}
  {1996})}\BibitemShut {NoStop}%
\bibitem [{\citenamefont {Hostler}(1970)}]{hostler1970}%
  \BibitemOpen
  \bibfield  {author} {\bibinfo {author} {\bibfnamefont {L.~C.}\ \bibnamefont
  {Hostler}},\ }\href@noop {} {\bibfield  {journal} {\bibinfo  {journal}
  {Journal of Mathematical Physics}\ }\textbf {\bibinfo {volume} {11}},\
  \bibinfo {pages} {2966} (\bibinfo {year} {1970})}\BibitemShut {NoStop}%
\bibitem [{\citenamefont {Maquet}\ \emph {et~al.}(1998)\citenamefont {Maquet},
  \citenamefont {V{\'e}niard},\ and\ \citenamefont {Marian}}]{Maquet1998}%
  \BibitemOpen
  \bibfield  {author} {\bibinfo {author} {\bibfnamefont {A.}~\bibnamefont
  {Maquet}}, \bibinfo {author} {\bibfnamefont {V.}~\bibnamefont
  {V{\'e}niard}},\ and\ \bibinfo {author} {\bibfnamefont {T.~A.}\ \bibnamefont
  {Marian}},\ }\href@noop {} {\bibfield  {journal} {\bibinfo  {journal}
  {Journal of Physics B: Atomic, Molecular and Optical Physics}\ }\textbf
  {\bibinfo {volume} {31}},\ \bibinfo {pages} {3743} (\bibinfo {year}
  {1998})}\BibitemShut {NoStop}%
\bibitem [{\citenamefont {Gradshteĭn}\ and\ \citenamefont
  {Ryzhik}(2015)}]{Gradshtein2015}%
  \BibitemOpen
  \bibfield  {author} {\bibinfo {author} {\bibfnamefont {I.~S.}\ \bibnamefont
  {Gradshteĭn}}\ and\ \bibinfo {author} {\bibfnamefont {I.~M.}\ \bibnamefont
  {Ryzhik}},\ }\href@noop {} {\emph {\bibinfo {title} {Table of integrals,
  series, and products}}},\ \bibinfo {edition} {8th}\ ed.\ (\bibinfo
  {publisher} {Academic Press},\ \bibinfo {year} {2015})\BibitemShut {NoStop}%
\bibitem [{\citenamefont {Szeg\"{o}}(1939)}]{Szego1939}%
  \BibitemOpen
  \bibfield  {author} {\bibinfo {author} {\bibfnamefont {G.}~\bibnamefont
  {Szeg\"{o}}},\ }\href@noop {} {\emph {\bibinfo {title} {Orthogonal
  Polynomials}}},\ \bibinfo {edition} {1st}\ ed.\ (\bibinfo  {publisher}
  {American Mathematical Society},\ \bibinfo {year} {1939})\BibitemShut
  {NoStop}%
\bibitem [{Jos()}]{Joseph2023code}%
  \BibitemOpen
  \href@noop {} {}\bibinfo {howpublished}
  {\url{https://github.com/JosephPScott/Hydrogen-S-state-2-photon}},\ \bibinfo
  {note} {accessed: 2023-0 7-18}\BibitemShut {NoStop}%
\bibitem [{\citenamefont {Marian}(1989)}]{Marian1989}%
  \BibitemOpen
  \bibfield  {author} {\bibinfo {author} {\bibfnamefont {T.~A.}\ \bibnamefont
  {Marian}},\ }\href@noop {} {\bibfield  {journal} {\bibinfo  {journal}
  {Physical Review A}\ }\textbf {\bibinfo {volume} {39}},\ \bibinfo {pages}
  {3816} (\bibinfo {year} {1989})}\BibitemShut {NoStop}%
\bibitem [{\citenamefont {Swainson}\ and\ \citenamefont
  {Drake}(1991)}]{Swainson1991}%
  \BibitemOpen
  \bibfield  {author} {\bibinfo {author} {\bibfnamefont {R.~A.}\ \bibnamefont
  {Swainson}}\ and\ \bibinfo {author} {\bibfnamefont {G.~W.}\ \bibnamefont
  {Drake}},\ }\href@noop {} {\bibfield  {journal} {\bibinfo  {journal} {Journal
  of Physics A: Mathematical and General}\ }\textbf {\bibinfo {volume} {24}},\
  \bibinfo {pages} {95} (\bibinfo {year} {1991})}\BibitemShut {NoStop}%
\bibitem [{\citenamefont {McNamara}\ \emph {et~al.}(2018)\citenamefont
  {McNamara}, \citenamefont {Fursa},\ and\ \citenamefont
  {Bray}}]{Mcnamara2018}%
  \BibitemOpen
  \bibfield  {author} {\bibinfo {author} {\bibfnamefont {K.}~\bibnamefont
  {McNamara}}, \bibinfo {author} {\bibfnamefont {D.}~\bibnamefont {Fursa}},\
  and\ \bibinfo {author} {\bibfnamefont {I.}~\bibnamefont {Bray}},\ }\href@noop
  {} {\bibfield  {journal} {\bibinfo  {journal} {Physical Review A}\ }\textbf
  {\bibinfo {volume} {98}},\ \bibinfo {pages} {043435} (\bibinfo {year}
  {2018})}\BibitemShut {NoStop}%
\bibitem [{\citenamefont {Singor}\ \emph {et~al.}(2021)\citenamefont {Singor},
  \citenamefont {Fursa}, \citenamefont {Bray},\ and\ \citenamefont
  {McEachran}}]{Singor2021}%
  \BibitemOpen
  \bibfield  {author} {\bibinfo {author} {\bibfnamefont {A.}~\bibnamefont
  {Singor}}, \bibinfo {author} {\bibfnamefont {D.}~\bibnamefont {Fursa}},
  \bibinfo {author} {\bibfnamefont {I.}~\bibnamefont {Bray}},\ and\ \bibinfo
  {author} {\bibfnamefont {R.}~\bibnamefont {McEachran}},\ }\href@noop {}
  {\bibfield  {journal} {\bibinfo  {journal} {Atoms}\ }\textbf {\bibinfo
  {volume} {9}},\ \bibinfo {pages} {42} (\bibinfo {year} {2021})}\BibitemShut
  {NoStop}%
\bibitem [{\citenamefont {Eschner}\ \emph {et~al.}(2003)\citenamefont
  {Eschner}, \citenamefont {Morigi}, \citenamefont {Schmidt-Kaler},\ and\
  \citenamefont {Blatt}}]{eschner2003}%
  \BibitemOpen
  \bibfield  {author} {\bibinfo {author} {\bibfnamefont {J.}~\bibnamefont
  {Eschner}}, \bibinfo {author} {\bibfnamefont {G.}~\bibnamefont {Morigi}},
  \bibinfo {author} {\bibfnamefont {F.}~\bibnamefont {Schmidt-Kaler}},\ and\
  \bibinfo {author} {\bibfnamefont {R.}~\bibnamefont {Blatt}},\ }\href@noop {}
  {\bibfield  {journal} {\bibinfo  {journal} {JOSA B}\ }\textbf {\bibinfo
  {volume} {20}},\ \bibinfo {pages} {1003} (\bibinfo {year}
  {2003})}\BibitemShut {NoStop}%
\end{thebibliography}%

\end{document}